\documentclass[11pt,draftcls,onecolumn]{IEEEtran}
\usepackage{ifpdf}
\ifpdf
	\usepackage[pdftex]{graphicx}
  \usepackage[update]{epstopdf}
\else
	\usepackage{graphicx}
\fi
\usepackage{times}
\usepackage{epsf}
\usepackage{epsfig}
\usepackage{amsmath}
\usepackage{amsfonts}
\usepackage{amssymb}
\usepackage{amstext}
\usepackage{latexsym}
\usepackage{color}
\usepackage{ifthen}
\usepackage{multirow}
\usepackage{subfigure}
\usepackage{cite,setspace}
\usepackage{algorithm}
\usepackage{algpseudocode}
\usepackage{setspace}

\newcommand{\bit}{\begin{itemize}}
\newcommand{\eit}{\end{itemize}}
\newcommand{\bcor}{\begin{cor}}
\newcommand{\ecor}{\end{cor}}
\newcommand{\beq}{\begin{equation}}
\newcommand{\eeq}{\end{equation}}
\newcommand{\beqn}{\begin{equation*}}
\newcommand{\eeqn}{\end{equation*}}
\newcommand{\bea}{\begin{eqnarray}}
\newcommand{\eea}{\end{eqnarray}}
\newcommand{\bean}{\begin{eqnarray*}}
\newcommand{\eean}{\end{eqnarray*}}
\newcommand{\ben}{\begin{enumerate}}
\newcommand{\een}{\end{enumerate}}
\newcommand{\bdefn}{\begin{defn}}
\newcommand{\edefn}{\end{defn}}
\newcommand{\bnote}{\begin{note}}
\newcommand{\enote}{\end{note}}
\newcommand{\bprop}{\begin{prop}}
\newcommand{\eprop}{\end{prop}}
\newcommand{\blem}{\begin{lem}}
\newcommand{\elem}{\end{lem}}
\newcommand{\bthm}{\begin{thm}}
\newcommand{\ethm}{\end{thm}}
\newcommand{\bconj}{\begin{conj}}
\newcommand{\econj}{\end{conj}}
\newcommand{\bconstr}{\begin{constr}}
\newcommand{\econstr}{\end{constr}}
\newcommand{\bpf}{\begin{IEEEproof}}
\newcommand{\epf}{\end{IEEEproof}}

\newcommand{\cmds}{\mbox{${\cal C}_{\text{\tiny MDS }}$}}
\newcommand{\gmds}{\mbox{$G_{\text{\tiny MDS}}$}}
\newcommand{\hmds}{\mbox{$H_{\text{\tiny MDS}}$}}

\newtheorem{theorem}{\textbf{Theorem}}
\newtheorem{prop}{\textbf{Proposition}}
\newtheorem{defn}[theorem]{Definition} 
\newtheorem{definition}{\textbf{Definition}}

\newtheorem{lemma}{\textbf{Lemma}}
\newtheorem{example}{Example}

\renewcommand{\vec}[1]{\mbox{\boldmath$#1$}}

\begin{document}

\title{Layered, Exact-Repair Regenerating Codes Via Embedded Error Correction and Block Designs}

\author{Chao Tian,~\IEEEmembership{Senior Member,~IEEE}, Birenjith Sasidharan, Vaneet Aggarwal,~\IEEEmembership{Member,~IEEE}, Vinay A. Vaishampayan,~\IEEEmembership{Fellow,~IEEE}, and  P. Vijay Kumar,~\IEEEmembership{Fellow,~IEEE}\thanks{This paper was presented in part at 2013 IEEE International Symposium on Information Theory, Istanbul, Turkey.}}

\maketitle

\begin{abstract}
A new class of exact-repair regenerating codes is constructed by stitching together shorter erasure correction codes, where the stitching pattern can be viewed as block designs. The proposed codes have the \lq\lq{}help-by-transfer\rq\rq{} property where the helper nodes simply transfer part of the stored data directly, without performing any computation. This embedded error correction structure makes the decoding process straightforward, and in some cases the complexity is very low. We show that this construction is able to achieve performance better than space-sharing between the minimum storage regenerating codes and the minimum repair-bandwidth regenerating codes, and it is the first class of codes to achieve this performance. In fact, it is shown that the proposed construction can achieve a non-trivial point on the optimal functional-repair tradeoff, and it is asymptotically optimal at high rate, {\em i.e.,} it asymptotically approaches the minimum storage and the minimum repair-bandwidth simultaneously.
\end{abstract}

\section{Introduction}

Distributed data storage systems can encode and disperse information (a message) to multiple storage nodes (or disks) such that a user can retrieve it by accessing only a subset of them. Such systems are able to provide superior reliability and availability in the event of disk corruption or network congestion. In order to reduce the amount of storage overhead required to guarantee such performance, erasure correction codes can be used instead of simple replication of the data. Given the massive amount of data that is currently being stored, even a small reduction in storage overhead can translate into huge savings. For instance, Facebook currently stores $3$ copies of all data, running $3000$ nodes with a total of $100$ PB of storage space. A $600$-node Hadoop \cite{hadoop} cluster at Facebook for performing data analytics on event logs from their website stores $2$ petabytes of data, and grows about $15$ TB every day~\cite{zaharia2010delay}.

When the data is encoded by an erasure code, data repair ({\em e.g.}, due to node failure) becomes more involved, because the information stored at a given node may not be directly available from any one of the remaining storage nodes, but it can nevertheless be reconstructed since it is a function of the information stored at these nodes. One key issue that affects the system performance is the total amount of information that the remaining nodes need to transmit to the new node. 
Consider a storage system which has $n$ storage nodes, and the data can be reconstructed by accessing any $k$ of them. A failed node is repaired by requesting any $d$ of the remaining nodes to provide information, and then using the received information to construct a new data storage node. A naive approach is to let these helper nodes transmit sufficient data such that the underlying data can be reconstructed completely, and then the information that needs to be stored at the new node can be subsequently generated. This approach is however rather wasteful, since the data stored at the new node is only a fraction of the complete data. 

Dimakis {\em et al.} in \cite{Dimakis:10} proposed the regenerating code framework to investigate the tradeoff between the amount of storage at each node ({\em i.e.,} data storage) and the amount of data transfer for repair ({\em i.e.,} repair bandwidth). It was shown that for the case when the regenerated new node only needs to fulfill the role of the failed node functionally ({\em i.e.}, functional-repair), but not to replicate exactly the original content at the failed node ({\em i.e.}, exact-repair), the problem can be converted to a network multicast problem, and thus the celebrated network coding result \cite{Yeung:00} can be applied.  By way of this equivalence, the optimal tradeoff was completely characterized in \cite{Dimakis:10} for this case. The two important extreme cases, where the data storage is minimized and the repair bandwidth is minimized, are referred to as minimum storage regenerating (MSR) codes and minimum bandwidth regenerating (MBR) codes, respectively.  The functional-repair problem is well understood and constructions of such codes are available (see \cite{Dimakis:10,Wu:10,Dimakis:11}).

The functional-repair framework implies that the coding rule evolves over time, which incurs additional system overhead. Furthermore, functional repair does not guarantee the data to be stored in systematic form, thus cannot satisfy this important practical requirement. In contrast, exact-repair regenerating codes do not suffer from such disadvantages. Exact-repair regenerating codes were investigated in \cite{RashmiShah:11,Tamo:11,RashmiShah:12:1,RashmiShah:12:2,Cadambe:11,PapailiopoulosDimakisCadambe:11,CadambeHuang:11}, all of which address either the MBR case or the MSR case. Particularly, the optimal code constructions in \cite{RashmiShah:12:1} and \cite{RashmiShah:11} show that the more stringent exact-repair requirement does not incur any penalty for the MBR case; the constructions in \cite{RashmiShah:12:2,RashmiShah:11,Cadambe:11} show that this is also true asymptotically for the MSR case. These results may lead to the impression that enforcing exact-repair never incurs any penalty compared to functional repair. However, it was shown in \cite{RashmiShah:12:1} that a large portion of the optimal tradeoffs achievable by functional-repair codes cannot be strictly achieved by exact-repair codes, and it was shown more recently in \cite{Tian:12} that there exists a non-vanishing gap between the optimal functional-repair tradeoff and the exact-repair tradeoff, and thus the loss is not asymptotically diminishing. The characterization of the optimal tradeoff for exact-repair regenerating codes under general set of parameters remains open. 

Codes achieving tradeoff other than the MBR point or the MSR point may be more suitable for systems employing exact-repair regenerating codes, which may have an acceptable storage-repair-bandwidth tradeoff and lower coding complexity. However, it is unknown whether there even exist codes that can achieve a storage-bandwidth tradeoff better than simply space-sharing between an MBR code and an MSR code.
In this work, we provide a code construction based on stitching together shorter erasure correction codes through combinatorial block designs, which is indeed able to achieve such tradeoff points. We show that it can achieve a non-trivial point on the optimal functional-repair tradeoff for $[n,k,d=k=n-1]$, and it is also asymptotically optimal at high rate while the space-sharing approach is strictly sub-optimal; moreover, space-sharing among this non-trivial tradeoff point, the MSR point, and the MBR point achieves the complete exact-repair tradeoff for the case $[n,k,d]=[4,3,3]$ given in \cite{Tian:12}.

The conceptually straightforward code construction we propose has the property that the helper nodes in the repair do not need to perform any computation, but can simply transmit certain stored information for the new node to synthesize and recover the lost information. This \lq\lq{}help-by-transfer\rq\rq{} property is appealing in practice, since it reduces and almost completely eliminates the computation burden at the helper nodes. This property also holds in the constructions proposed in \cite{RashmiShah:12:1} and \cite{Papailiopoulos:12INFOCOM}. In fact our construction was partially inspired by and may be viewed as a generalization of these codes. Another closely related work is \cite{Rouayheb:10}, where block designs were also used, however repetition is the main tool used in that construction, in contrast to the embedded erasure correction codes in our construction. The system model in \cite{Rouayheb:10} is also different, where the repair only needs to guarantee the existence of one particular $d$-helper-node combination (fix-access repair), instead of the more stringent requirement that the repair information can come from any $d$-helper-node combination (random-access repair). 


The results presented here are the combination of two independent and concurrent works \cite{Tian:Arxiv} and \cite{Birenjith:13}. Given the surprising similarity between the code constructions found by the two groups, we decided to merge the results in the hope that the readers may gain a more coherent understanding from this effort\footnote{It should be noted that the \lq\lq{}layers\rq\rq{} in \cite{Tian:Arxiv} and \cite{Birenjith:13} refer to different aspects of the construction: in the former it is used to refer the concatenation of two erasure correction coding steps, while in the latter it is used to refer to the way the component codes are arranged.}.

The rest of the paper is organized as follows. In Section \ref{sec:definition}, several relevant existing results are reviewed. Section \ref{sec:canonical} provides the construction of the canonical codes for the case $[n,k,d=k]$, and the performance is analyzed. Section \ref{sec:general} provides the general code constructions. Finally Section \ref{sec:conclusion} concludes the paper.

\section{Preliminaries}
\label{sec:definition}

In this section, we review some basics on regenerating codes, maximum separable codes, rank metric codes, and combinatorial block designs. We write $\{1,2,\ldots,n\}$ as $I_n$ for simplicity.

\subsection{Exact-Repair Regenerating Codes}

An $[n,k,d]$ exact-repair regenerating code for a storage system with a total of $n$ storage nodes satisfies the condition that any $k$ of them can be used to reconstruct the original message, and to repair a lost node, the new node may access data from any $d$ of the remaining nodes. Let the total amount of raw data stored be $M$ units and let each storage site store $\alpha$ units, {\em i.e.,} the redundancy of the system is $n\alpha-M$. To repair a node failure (regenerate a new node), each helper node transmits $\beta$ units of data to the new node, which results in a total of $d\beta$ units of data transfer. It is clear that the quantities $\alpha$ and $\beta$ scale linearly with $M$, and thus we shall normalize the other two quantities using $M$, {\em i.e.,}
\begin{align}
\bar{\alpha}\triangleq \frac{\alpha}{M},\quad \bar{\beta}\triangleq\frac{\beta}{M}, 
\end{align}
and use them as the measure of performance from here on. The problem can be more formally defined using a set of encoding and decoding functions, which we omit here for conciseness (see \cite{Tian:12}).  
 
\subsection{Cut-Set Outer Bound, the MBR Point and the MSR Point}
\begin{figure}[tcb]
  \centering
  \includegraphics[width=8cm]{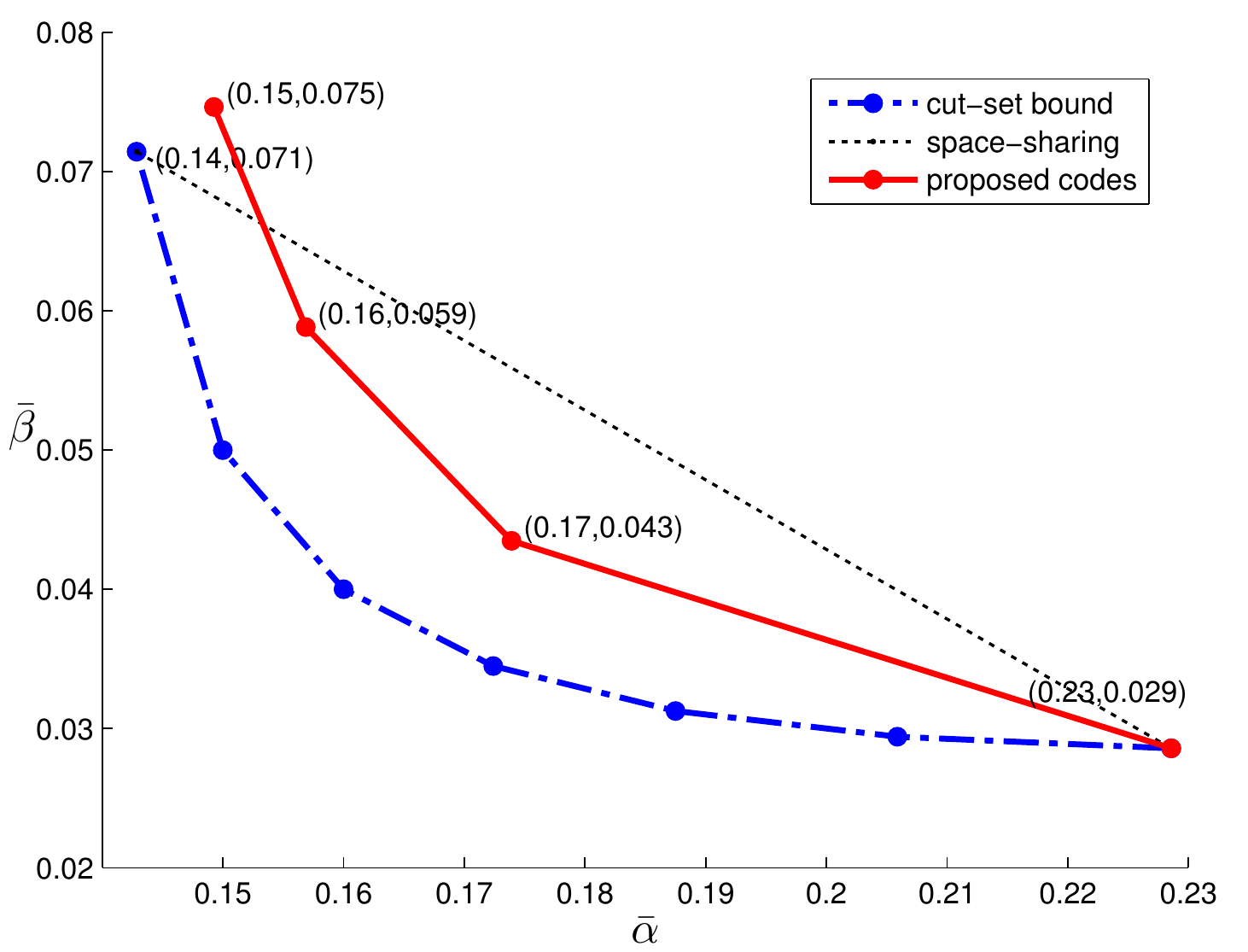}
  \caption{The cut-set bound, the space-sharing line and the tradeoffs achieved by the proposed codes for $[n,k,d]=[9,7,8]$.\label{fig:cutset}}
\end{figure}

As mentioned earlier, a precise characterization of the optimal storage-bandwidth tradeoff under functional repair was obtained in \cite{Dimakis:10}, which is given by
\begin{align}
\sum_{i=0}^{k-1}\min(\bar{\alpha},(d-i)\bar{\beta})\geq 1. \label{eqn:cutsettradeoff}
\end{align} 
Since exact-repair is a more stringent requirement than functional-repair, it provides an outer bound for exact-repair regenerating codes, which must also satisfy (\ref{eqn:cutsettradeoff}), possibly with strict inequality. It can be shown that the bound in (\ref{eqn:cutsettradeoff}) is equivalently to
\begin{align}
p\bar{\alpha}+\sum_{i=p}^{k-1}(d-i)\bar{\beta}\geq 1,\qquad p=0,1,\ldots,k-1. \label{eqn:cutsettradeoffequivalent}
\end{align}

One extreme point of this outer bound is when the storage is minimized, {\em i.e.}, the minimum storage regenerating (MSR) point, which is
\begin{align}
\bar{\alpha}=\frac{1}{k}, \quad \bar{\beta}=\frac{1}{k(d-k+1)}.
\end{align}
The other extreme case is when the repair bandwidth is minimized, {\em i.e.}, the minimum bandwidth regenerating (MBR) point, which is
\begin{align}
\bar{\alpha} =\frac{2d}{k(2d-k+1)},\quad\bar{\beta}=\frac{2}{k(2d-k+1)}.
\end{align}
Both of these extreme points (on the functional-repair tradefoff) are achievable (see \cite{RashmiShah:11,RashmiShah:12:1,RashmiShah:12:2,Cadambe:11}) under exact-repair, however the functional-repair outer bound is not tight in general (see \cite{RashmiShah:12:1} and \cite{Tian:12}). The outer bound and the two extreme points are illustrated in Fig. \ref{fig:cutset} for $[n,k,d]=[9,7,8]$.

The space-sharing line between MSR and MBR points is characterized by the equation ({\em e.g.,}\cite{RashmiShah:12:1})
\bea 
k\bar{\alpha}+k(d-k+1)\bar{\beta} & = & 2,\label{eq:space_sharing}
\eea
which when $d=k$, reduces to
\bea \label{eq:space_sharing_k}
k(\bar{\alpha}+\bar{\beta}) & = & 2.
\eea

It is sometimes convenient to view all the achievable $(\bar{\alpha},\bar{\beta})$ pairs together as a region, for which we introduce the following definition. 
\begin{definition}
A pair $(\bar{\alpha},\bar{\beta})$ is said to be achievable for $[n,k,d]$ exact-repair regenerating if there exists an exact-repair regenerating code with such a normalized storage and repair-bandwidth. The closure of the collection of all such pairs is the achievable $(\bar{\alpha},\bar{\beta})$ region, denoted as $\mathcal{R}_{n,k,d}$.   
\end{definition}

\subsection{Asymptotic Tradeoff Region}

The proposed codes have performance better than space-sharing line in many cases, especially when $k$ is close to $n$. It is insightful to consider the asymptote when $k$ is driven to infinity while keeping $n=k+\tau_1$ and $d=k+\tau_2$ where $\tau_1$ and $\tau_2$ are fixed constant integers such that $\tau_1>\tau_2\geq 0$. For this purpose, define the following region 
\begin{align}
\mathcal{R}_{\infty}\triangleq\bigcup_{k\rightarrow \infty} k\mathcal{R}_{(k+\tau_1,k,k+\tau_2)},
\end{align}
where $\tau_1$ and $\tau_2$ are fixed integers as previously stated, and we have multiplied the components of elements in $\mathcal{R}_{(k+\tau_1,k,k+\tau_2)}$ by $k$. This $k$-fold expansion definition is partly motivated by observing $k$ appears for both $\bar{\alpha}$ and  $\bar{\beta}$ terms in (\ref{eq:space_sharing}).

It is trivial to see that an outer bound for $\mathcal{R}_{\infty}$  is given by 
\begin{align}
k\bar{\alpha}\geq 1,\quad k\bar{\beta}\geq 0,
\label{eqn:outerboundR}
\end{align}
by taking $\bar{\alpha}$ at the MSR point, and $\bar{\beta}$ at the MBR point.

Space-sharing between the MSR point and the MBR point cannot achieve this outer bound due to (\ref{eq:space_sharing}). In Section \ref{sec:canonical}, we show that the proposed codes can achieve the entire region $\mathcal{R}_{\infty}$ when $d=k$.

\subsection{Maximum Distance Separable Code}

A linear code of length-$n$ and dimension $k$ is called an $[n,k]$ code. The Singleton bound (see {\em e.g.}, \cite{Wicker:book}) is a well known upper bound on the minimum distance for any $[n,k]$ code, given as
\begin{align}
d_{\min}\leq n-k+1.
\end{align}

An $[n,k]$ code that satisfies the Singleton bound with equality is called a maximum distance separable (MDS) code. A key property of an MDS code is that it can correct any $(n-k)$ or fewer erasures. There exist various ways to construct MDS codes for any given $[n,k]$ values, and it is known that there exists an $[n,k]$ MDS code in any finite field $\mathbb{F}_q$  where $q\geq n$; see, {\em e.g.}, \cite{Wicker:book}. 

In coding literature, an $[n,k]$ code with minimum distance $d_{\min}$ is sometimes also referred to as an $(n,k,d_{\min})$ code. In the context of regenerating codes, the triple $[n,k,d]$ instead specifies the total number of nodes, the number of nodes that together allow reconstruction of the data, and the number of helper nodes during a repair, respectively. In order to avoid possible confusion, we do not write the minimum distance $d_{\min}$ explicitly for a linear code, and also use brackets instead of parentheses in this work. 

\subsection{Linearized Polynomial and Gabidulin Codes}

An important component in our construction is a code based on linearized polynomials, and the following lemma is particularly relevant to us; see, {\em e.g.,} \cite{LidNie}.

\begin{lemma} \label{lem:useful}
A linearized polynomial 
\bean
f(x) & = & \sum_{i=1}^{M}v_i x^{q^{i-1}}, \  v_i \in \mathbb{F}_{q^\kappa}
\eean
can be uniquely identified from evaluations at any $M$ points, for which the input values are linearly independent 
over $\mathbb{F}_q$. 
\end{lemma}

Another relevant property of linear polynomials is that they satisfy the following condition 
\bean
f(ax + by) & = & af(x) + bf(y), \ a, b \in \mathbb{F}_q, \ x, y \in \mathbb{F}_{q^\kappa},
\eean
which is the reason that they are called \lq\lq{}linearized\rq\rq{}.

Gabidulin \cite{Gab85} proposed a class of codes based on linearized polynomials, which is maximum distance separable in terms of rank metric. This class of codes can be viewed as a generalized version of the MDS codes, and it plays an instrumental role in our construction.  

\subsection{Block Designs}

\begin{table}[t]
\caption{Example Steiner systems $S(2,3,7)$, $S(2,3,9)$ and $S(2,4,13)$.}
\label{table:steiner79}
\centering
\begin{tabular}{|c|c|}
\hline
$ S(3,7)$&$\{(1,2,3),(1,4,5),(1,6,7),(2,4,6),(2,5,7),(3,4,7),(3,5,6)\}$\\\hline
$S(3,9)$&$\{(2,3,4),(5,6,7),(1,8,9),(1,4,7),(1,3,5),(4,6,8),$\\
&$(2,7,9),(2,5,8),(1,2,6),(4,5,9),(3,7,8),(3,6,9)\}$\\\hline
$S(4,13)$&$\{(1,2,4,10),(2,3,5,11),(3,4,6,12),(4,5,7,13),(5,6,8,1),$\\
&$(6,7,9,2),(7,8,10,3),(8,9,11,4),(9,10,12,5),$\\
&$(10,11,13,6),(11,12,1,7),(12,13,2,8),(13,1,3,9)\}$\\\hline
\end{tabular}
\end{table}

\begin{table}
\caption{$\gamma$ and $N$ values for the two classes of block designs.}
\label{table:gammaN}
\centering
\begin{tabular}{|c|c|c|}
\hline
& $\gamma$& $N$\\\hline
DCBD&$\nu {n-1 \choose r-1}$&$\nu{n \choose r}$\\
\hline
BIBD&$\frac{\lambda(n-1)}{r-1}$&$\frac{\lambda n(n-1)}{r(r-1)}$\\
\hline
\end{tabular}
\end{table}

A block design is a set together with a family of subsets ({\em i.e.}, blocks) whose members are chosen to satisfy some properties. The blocks are required to all have the same number of elements, and thus a given block design with parameters $(r,n)$, where $r<n$, is specified by $(X,\mathcal{B})$ where $X$ is an $n$-element set and $\mathcal{B}$ is a collection of $r$-element subsets of $X$. The blocks are usually allowed to repeat. 
We use $N$ to denote the total number of blocks in a block design when the parameters are clear from the context. 
Two classes of block designs are particularly relevant to us:
\begin{itemize}
\item The first is a restricted class of Steiner systems known in the literature. A Steiner system $S(t,r,n)$ is a block design with parameters $(r,n)$ where each element of $X$ appears exactly $\gamma$ times, and each $t$-element subset of $X$ appears in exactly one block; in this work we shall restrict our attention to the case $t=2$, and thus refer to it as a restricted Steiner system and write it simply as $S(r,n)$. This design can be generalized to balanced incomplete block design (BIBD), $S_\lambda(r,n)$, where each pair of elements of $X$ appears in exactly $\lambda$ blocks, instead of a single block. A restricted Steiner system is thus a BIBD with $\lambda=1$.
\item We refer the second class of block designs as duplicated combination block design (DCBD). An $r$-combination of a set $X$ is a subset of $r$ distinct elements of $X$. A duplicated combination block design  $C_\nu(r,n)$ is a block design with parameters $(r,n)$ where each $r$-combination appears exactly $\nu$ times, which we write as $C_\nu(r,n)$.
\end{itemize}

It is clear that DCBDs can be viewed as BIBDs with
$\lambda = \nu{n-2 \choose r-2}$.
This implies that for any $(r,n)$ pair, a BIBD always exists (in fact even when we limit to $\nu=1$). However, for a fixed $(\lambda,r)$ pair, a BIBD may not exist for all values of $n$. For the particularly 
well understood Steiner triple systems ({\em i.e.} Steiner systems when $t=2$ and $r=3$), there exists an $S(3,n)$ if and only if $n=0$, or $n$ modulo $6$ is $1$ or $3$  \cite{Colbourn:book}. Examples of $S(3,7)$, $S(3,9)$ are given in Table \ref{table:steiner79}, where a design for $S(4,13)$ is also included. The parameter $\gamma$ and the total number of blocks $N$ can be calculated straightforwardly (see \cite{Colbourn:book}), and are listed in Table \ref{table:gammaN} for convenience.   Without loss of generality, we assume $X=I_n$ from here on. 
More details on BIBDs, Steiner systems and other block designs can be found in, {\em e.g.}, \cite{Colbourn:book} and  \cite{Bose:39}.

\section{Canonical Codes for $[n,k,d=k]$}
\label{sec:canonical}

In this section, we present a set of exact-repair codes, referred to as the canonical codes, for the case $d=k$. The overall code is formed by stitching together shorter MDS codes, and the stitching patterns follow either BIBDs or DCBDs. This set of codes can be indexed by two auxiliary parameters $m$ and $r$ satisfying $1\leq m< r< n$, where $r$ is the same parameter as in the block designs being used. As will be seen, the parameters $d$ and $m$ are related as $m=n-d$, and the codes for $m=1$ are particularly simple which will be presented first. The qualifier ``canonical'' is used to describe the case of $d=k$ because the construction in this case can be viewed as the basic form of a subsequent construction for the general case $k < d$.

The canonical codes, together with known MSR codes and MBR codes, achieve the complete optimal tradeoff for $[n,k,d]=[4,3,3]$ that was recently characterized in \cite{Tian:12}. For $[n,k,d=k=n-1]$, this construction is always able to achieve a non-trivial point on the cut-set bound, {\em i.e.,} the optimal functional-repair tradeoff, other than the MSR point and the MBR point.  More generally, for $[n,k,d=k]$, it can achieve performance better than space-sharing between MSR and MBR in certain parameter range. For high rate regenerating codes, the canonical codes are asymptotically optimal, and essentially achieve the complete region $\mathcal{R}_\infty$.

\subsection{Canonical Codes Using Restricted Steiner Systems and BIBDs}

\begin{figure}[tcb]
  \centering
  \includegraphics[width=16cm]{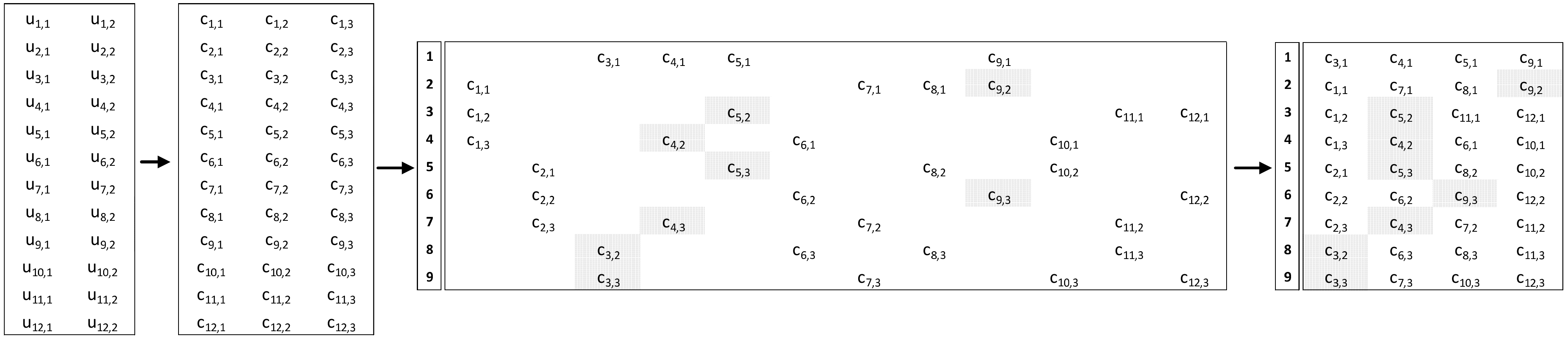}
  \caption{\label{fig:988example}  For $[n,k,d]=[9,8,8]$, the parameter chosen here are $m=1$, $r=3$, and the block design is Steiner triple system shown on the second row of Table \ref{table:steiner79}. The data matrix is of dimension $12\times 2$, and after the first encoding step, $\vec{c_i}=(u_{i,1},u_{i,2},u_{i,1}+u_{i,2})$ are placed on the $i$-th column in the auxiliary matrix (the third matrix). The resulting code matrix is of dimension $9\times 4$, after the blank spaces are removed. The helper symbols to repair node-$1$ are given in shade. }
\end{figure}

We use restricted Steiner Systems and BIBDs to construct canonical codes for the cases $d=k=n-1$. Here the auxiliary parameter $m=1$, and it will become clear in the sequel why it is set as such. First fix a restricted Steiner system $S(r,n)=\{B_1,B_2,\ldots,B_N\}$. The canonical code using this block design has $M_c=(r-m)N=(r-1)N$ data symbols in certain finite field $\mathbb{F}_q$, arranged as an $N\times(r-1)$ matrix $\vec{U}$, whose rows are $\vec{u_1},\vec{u_2},\ldots,\vec{u_N}$. The structure of the canonical code can be inferred from a two-step process (see Fig. \ref{fig:988example}) by which the data matrix $\vec{U}$ is encoded into an $n\times \gamma$ code array:
\ben
\item For $i=1,2,...,N$, the vector $\vec{u_i}$ is encoded into $\vec{c_i}=(u_{i,1},u_{i,2},\ldots,u_{i,r-1},\sum_{j=1}^{r-1}u_{i,j})$;
\item The $r$ symbols in $\vec{c_i}$, referred to together as a parity group, are placed in the rows specified in $B_i$, $i=1,2,\ldots,N$, appended after any previous written symbols\footnote{All the symbols in $\vec{c_i}$ together are sometimes called a parity group in the storage literature, and are referred to as a layer in \cite{Birenjith:13}; we shall adopt the parity group terminology in the sequel.}. 
\een

After these encoding steps, each row in the resulting matrix corresponds to the symbols to be written on each node. Since the arrangement of the blocks is not unique, and the placement of the symbols in each parity group $\vec{c_i}$ is also not unique, consequently the resulting code is not unique. 

Since each component code $\vec{c_i}$ has one parity symbol, it can withstand up to one erasure ($m=1$), and thus any single lost node can be repaired from the other $n-1$ nodes. More precisely, to repair node $j$, the helper node set is $\Delta=I_n\setminus \{j\}$ and the repair process has two steps  (see Fig. \ref{fig:988example}):
\ben
\item Helper transmission: For $i=1,2,...,N$, if $j\in B_i$, then the helper nodes in $\Delta\cap B_i$  ({\em i.e.,} the helper nodes that have symbols in $\vec{c_i}$) send the symbols in $\vec{c_i}$ to the new node;
\item Symbol regeneration:  For $i=1,2,...,N$, if $j\in B_i$, with the $r-1$ symbols received from the helper nodes,  the lost symbol in $\vec{c_i}$ is regenerated.
\een
Based on the construction, it can be seen that 
\begin{align}
M_c=(r-1)N=\frac{ n (n-1)}{r},\quad \alpha=\gamma=\frac{(n-1)}{r-1},\qquad \beta=1,
\end{align}
where the value of $\alpha$ is derived from the fact that in restricted Steiner systems each element appears in exactly $\gamma$ blocks, and the value of $\beta$ is derived from the fact that node $j$ contributes one symbol to repair node $i$ whenever $(i,j)$ appears in a block in the block design, and the fact that each pair of elements appears in exactly one block. Clearly the alphabet here can be chosen as $\mathbb{F}_2$, {\em i.e,}, a binary code. 

In the construction, the restricted Steiner system can be replaced with a more general BIBD without any essential change, resulting in the parameters
\begin{align}
M_c=(r-1)N=\frac{\lambda n (n-1)}{r},\quad \alpha=\gamma=\frac{\lambda(n-1)}{r-1},\qquad \beta=\lambda.
\end{align}

\subsection{Canonical Codes Using DCBDs}

As a natural generalization from the previous case, for $d=k\leq n-1$ we set the auxiliary parameter $m=n-d$. Intuitively $m$ is again the number of erasures that the component codes $\vec{c_i}$ can withstand, and since having $d=n-m$ helper nodes can be equivalently viewed as erasing the other $m$ nodes, any lost symbols can be regenerated using only $d=n-m$ helper nodes. For the repetition factor $\nu$, let us for now choose $\nu=d=n-m$, and we will revisit it later to discuss possibly reducing its value. 

\begin{figure}[tcb]
  \centering
  \includegraphics[width=16cm]{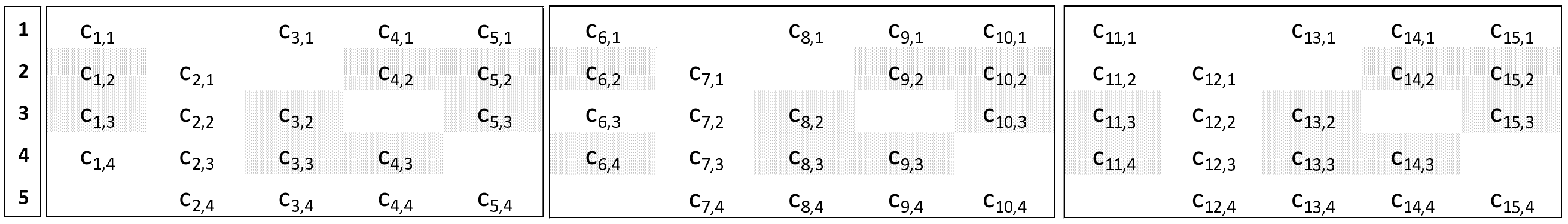}
  \caption{\label{fig:533example} For $[n,k,d]=[5,3,3]$, the parameter is chosen as $m=n-d=2$, $r=4$, and the block design is $3$-DCBD with parameters $(4,5)$, which duplicate the following blocks three times: $\{(1,2,3,4),(2,3,4,5),(1,3,4,5),(1,2,4,5),(1,2,3,5)\}$. Only the auxiliary form in encoding step (2) is shown here. The data matrix is of size $15\times2 $ and the resulting code matrix is of $5\times 12$ (after removing the blank spaces).  The helper symbols to repair node-$1$ are highlighted.}
\end{figure}

Fix a $C_\nu(r,n)=\{B_1,B_2,\ldots,B_N\}$. We encode an $N\times(r-m)$ matrix into an $n\times \gamma$ code array in two steps (see Fig. \ref{fig:533example}):
\ben
\item For $i=1,2,...,N$, the vector $\vec{u_i}$ is encoded using an $[r,r-m]$ MDS code to yield $\vec{c_i}$
\bean
\vec{u}_i \in \mathbb{F}_q^{r-m} & \Rightarrow & \vec{c}_i \in \mathbb{F}_q^{r}.
\eean
\item The $r$ symbols in $\vec{c_i}$ are placed in the rows specified in $B_i\in C_{\nu}(r,n)$, appended after any previous written symbols. 
\een

The only difference from the previous case is that the encoding from $\vec{u_i}$ into $\vec{c_i}$ now utilizes a general MDS code, instead of the single parity code (also an MDS code). The alphabet here can be chosen to be any $\mathbb{F}_q$ where $q\geq r$, in order for the component MDS code to exist. 
To repair node $j$, the helper node set is denoted as $\Delta=\{\delta_1,\delta_2,\ldots,\delta_d\}$, and the repair process is as follows  (Fig. \ref{fig:533example}):
\ben
\item Helper transmission: For $i=1,2,...,N$, if $j\in B_i$, some $(r-m)$ helper nodes in the set $\Delta\cap B_i$ send the symbols in $\vec{c_i}$ to the new node;
\item Symbol regeneration:  For $i=1,2,...,N$, if $j\in B_i$, with the $(r-m)$ symbols received from the helper nodes,  the lost symbol in $\vec{c_i}$ is regenerated.
\een

The choice of $m=n-d$ guarantees that the condition $|\Delta\cap B_i|\geq r-m$ holds as long as $|\Delta|=d\geq r-m$, thus the repair will always succeed. However, it may occur that $|\Delta\cap B_i|>r-m$ for some cases, {\em i.e.,} there may be more than one arrangement as to which $(r-m)$ helper nodes should transmit the symbols to regenerate the lost symbol in $\vec{c_i}$ ({\em e.g.,} in the first column of Fig. \ref{fig:533example} we can also choose $c_{1,2}$ and $c_{1,4}$ to repair $c_{1,1}$). Some combinations of the arrangements may result in transmissions being non-uniform among the helper nodes during repair. If we were to choose $\nu=1$, the resulting code can still repair a lost node however with non-uniform repair transmissions from the $d$ helper nodes, resulting in repair transmissions in the amounts of $\vec{\beta}=(\beta_1,\beta_2,\ldots,\beta_{d})$; it is clear that by using $\nu=d=n-m$, the code symbols in the duplicate portions can be repaired with transmission amounts which are circularly shifted versions of $\vec{\beta}$, and thus the total repair transmission amounts are uniform (see Fig. \ref{fig:533example}). In fact, the value of $\nu$ may be further reduced in some cases, as given in the following proposition whose proof can be found in the appendix. 

\begin{prop}
\label{prop:repetition}
For every integer $p$, $1 \leq p \leq m<r$, define
\bean
\theta_p = (d,r-p)_{\text{gcd}}, \qquad \zeta_p  =  \text{lcm}\left\{ \frac{\left(\frac{\theta_p}{s}\right)}{ \left(\left(\frac{\theta_p}{s}\right),r-m\right)_{\text{gcd}}}: s \mid \theta_p \right\}, \qquad \eta_p  =  \frac{\zeta_p}{(\zeta_p,{m-1 \choose p-1})_{\text{gcd}}} ,
\eean
where $(a,b)_{\text{gcd}}$ is the greatest common divisor of positive integers $a$ and $b$, and $a\mid b$ means $a$ is divides $b$. 
Then, $\nu$ can be set as
\bean 
\nu & = & \text{lcm} \{ \eta_p \mid 1 \leq p \leq m \},
\eean
and there exists a repair pattern such that the transmissions are uniform among all the $d$ helpers.  
\end{prop}

Note that $\nu$ is always a factor of $d$. Whenever $d$ is a prime with $r \leq d$, it can be checked that $\nu=1$. Even when it is not, $\nu$ can become $1$ in many cases. For example, when $d=8, r = 6, m=2$, it can be checked that $\nu = 1$.  

It is clear from the above discussion that
\begin{align}
M_c=(r-m)N=(r-m)\nu{n \choose r},\quad\alpha=\gamma=\nu{n-1 \choose r-1},\quad \beta=\frac{(r-m)\alpha}{n-m}=\frac{(r-m)\nu}{n-m}{n-1 \choose r-1},
\end{align}
where $\beta$ is derived from the total amount of repair transmission and the fact it can be distributed uniformly among the $d$ helper nodes. 

For the case $d=k=n-1$, DCBDs with parameters $(r,n)$ can also be used to construct canonical codes even when restricted Steiner systems $S(r,n)$ (or BIBDs $S_{\lambda}(r,n)$) indeed exist; it can be verified that such constructions in fact does not change the resultant $(\bar{\alpha},\bar{\beta})$. The advantage of using restricted Steiner systems and BIBDs is that the codes have smaller $\alpha$ and $\beta$ values, and thus practically more versatile. For example, the code in Fig. \ref{fig:988example} has $\alpha=4$ and $\beta=1$; on the other hand, the corresponding code using DCBDs in the same alphabet has $\alpha=28$ and $\beta=7$.

It should also be noted that for the case $d<n-1$, we can utilize general Steiner systems ({\em i.e.,} when $t>2$) or a more general class of block designs called $t$-designs, to construct canonical codes. However, the problem of non-uniform repair transmissions becomes rather intractable. Moreover, it was shown in \cite{Tian:Arxiv} that the non-canonical codes based on such constructions may induce loss of performance in terms of the normalized storage-repair-bandwidth tradeoff, when compared to that based on DCBDs unless certain additional conditions are met (more precisely, the uniform-rank-accumulation property given in Section \ref{sec:ura}). We thus do not pursue this route further. 

\subsection{Performance Assessment of Canonical Codes} 
\label{sec:can_performance}

We next state several results pertaining to the performance of the canonical code. The first result characterizes the range of the auxiliary parameters $(r,m)$ for which canonical codes outperform space-sharing between MSR and MBR points. Then we show that the canonical construction yields optimal codes operating on the functional-repair tradeoff when $d=k=n-1$. The third result is regarding the asymptotic optimality of the canonical codes at high rates. 

For canonical codes using DCBDs, the normalized storage and repair bandwidth $(\bar{\alpha},\bar{\beta})$ pair is 
\begin{align}
(\bar{\alpha},\bar{\beta})=\left(\frac{r}{n(r-m)},\frac{r}{n(n-m)}\right),\label{eqn:achievablealphabeta}
\end{align} 
and it can be verified that taking $m=1$ reduces (\ref{eqn:achievablealphabeta}) to that induced by codes based on BIBDs.

\begin{prop} 
\label{prop:optimal}
The $[n,k,d=k]$-canonical code operates at an $(\bar{\alpha}, \bar{\beta)}$-point that lies in between the MSR and MBR points, and improves upon space-sharing between the MSR and MBR points, whenever $m < r-m < k$. 
\end{prop}
\bpf
 Substituting (\ref{eqn:achievablealphabeta}) into the left hand side of (\ref{eq:space_sharing_k}), the performance is better than space-sharing as long as 
\begin{align}
\left(\frac{kr}{n(r-m)}+\frac{r}{n}\right)<2,
\end{align}
which is equivalent to $r> 2m$ and $n>r$, and further equivalent to $k>r-m>m$, under which the performance of the canonical codes is strictly superior to space-sharing between MSR and MBR points.
\epf

Whenever $n < 2k-1$, there exists an $(r,m)$ choice to satisfy the condition given above, consequently an $[n,k,d=k]$-canonical code that performs better than space-sharing between MSR and MBR points. Conversely, when  $n \geq 2k-1$, such choice of $(r,m)$ does not exist, and thus the canonical codes do not provide any gain over the space-sharing approach.

\begin{prop} The $[n,k,d=k=n-1]$-canonical code can achieve $(\bar{\alpha},\bar{\beta})=\left(\frac{n-1}{n(n-2)},\frac{1}{n}\right)$, which is on the functional-repair tradeoff but not the MSR point or the MBR point.
\end{prop}
\bpf 
Choose $m=1$ and $r=n-1$ in (\ref{eqn:achievablealphabeta}) gives the normalized $(\bar{\alpha},\bar{\beta})$ pair specified above. Setting $p=k-1$ in the left hand side of (\ref{eqn:cutsettradeoffequivalent}), and substituting the above $(\bar{\alpha},\bar{\beta})$ pair, we have, 
\begin{align}
(k-1)\bar{\alpha}+\bar{\beta}=(k-1)\frac{n-1}{n(n-2)}+\frac{1}{n}=1,
\end{align}
{\em i.e.,} it lies on the cut-set bound, however it is not the MSR or the MBR points.
\epf

\begin{figure}[tcb]
  \centering
  \includegraphics[width=12cm]{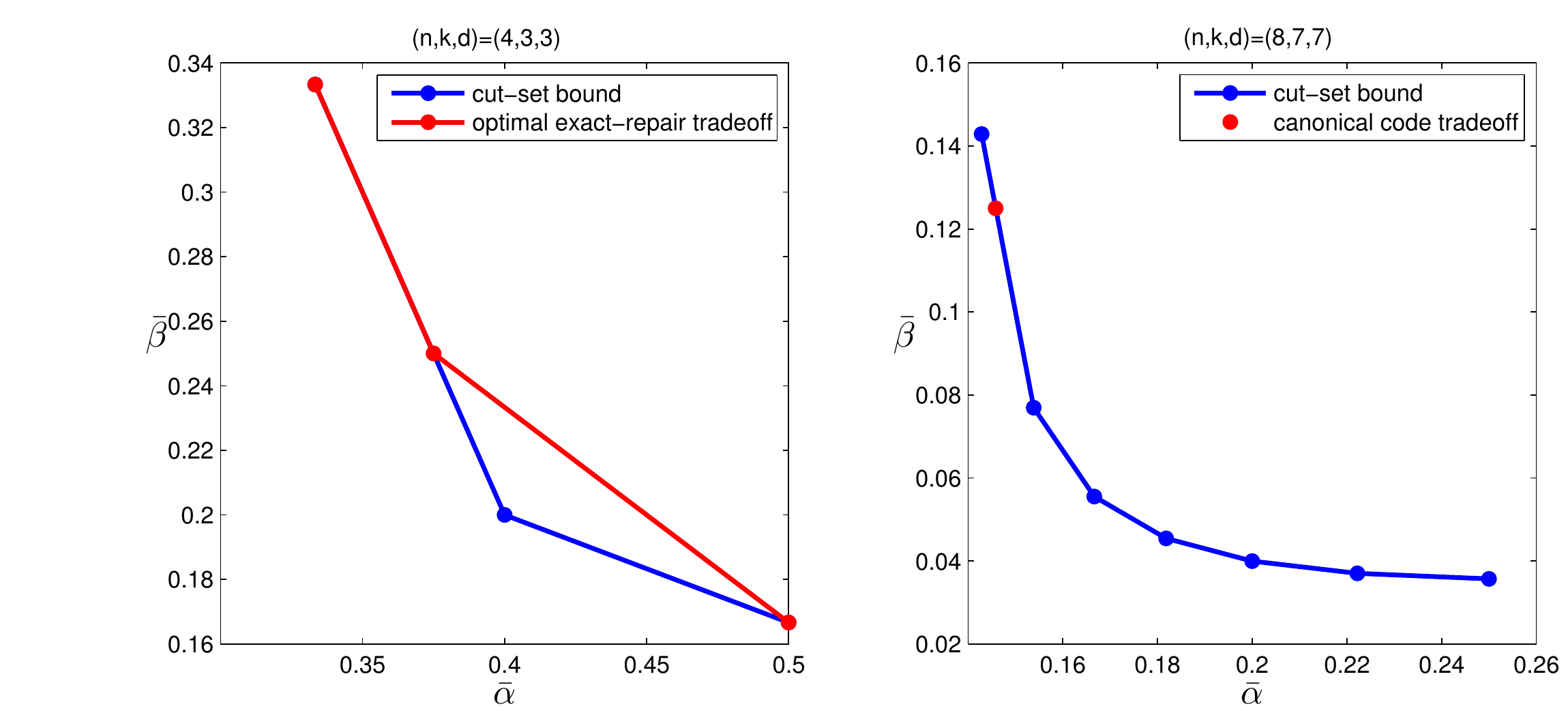}
  \caption{The tradeoff points of the canonical codes for $[n,k,d]=[4,3,3]$ and $[n,k,d]=[8,7,7]$ that are on the cut-set bound.\label{fig:optimalexample}}
\end{figure}

In Fig. \ref{fig:optimalexample}, two example cases of the tradeoff points achieved in Proposition \ref{prop:optimal} are given. For the particular case of $[n,k,d]=[4,3,3]$, space-sharing between the MSR point, the point achieved by the canonical code, and the MBR point characterizes the optimal exact-repair tradeoff, which was dervied in \cite{Tian:12}. The non-achievability result established in \cite{RashmiShah:12:1} does not apply to a narrow line-segment close to the MSR point, and the point given the above lemma indeed lies in this region. 

\begin{prop}
The region $\mathcal{R}_\infty$ is given by the set of pairs satisfying (\ref{eqn:outerboundR}), and it can be achieved using the canonical codes when $d=k$.
\end{prop}
\begin{IEEEproof}
We show that the canonical codes can achieve asymptotically 
\begin{align}
k\bar{\alpha}\rightarrow 1,\quad k\bar{\beta}\rightarrow0,
\end{align}
which is the only non-trivial corner point of the the outer bound region given in (\ref{eqn:outerboundR}).

Notice that by choosing $r=\sqrt{k}$ and $m=n-d=\tau_1-\tau_2=\tau_1$ in this case, we have
\begin{align}
\lim_{k\rightarrow \infty}k\frac{r}{n(r-m)}= \lim_{k\rightarrow \infty}\frac{k\sqrt{k}}{(k+\tau_1)(\sqrt{k}-\tau_1)}=1,
\end{align}
and 
\begin{align}
\lim_{k\rightarrow \infty}k\frac{r}{n(n-m)}=\lim_{k\rightarrow \infty}\frac{k\sqrt{k}}{(k+\tau_1)k}=0.
\end{align}
The proof is thus complete.
\end{IEEEproof}

\begin{figure}[tcb]
  \centering
  \includegraphics[width=8cm]{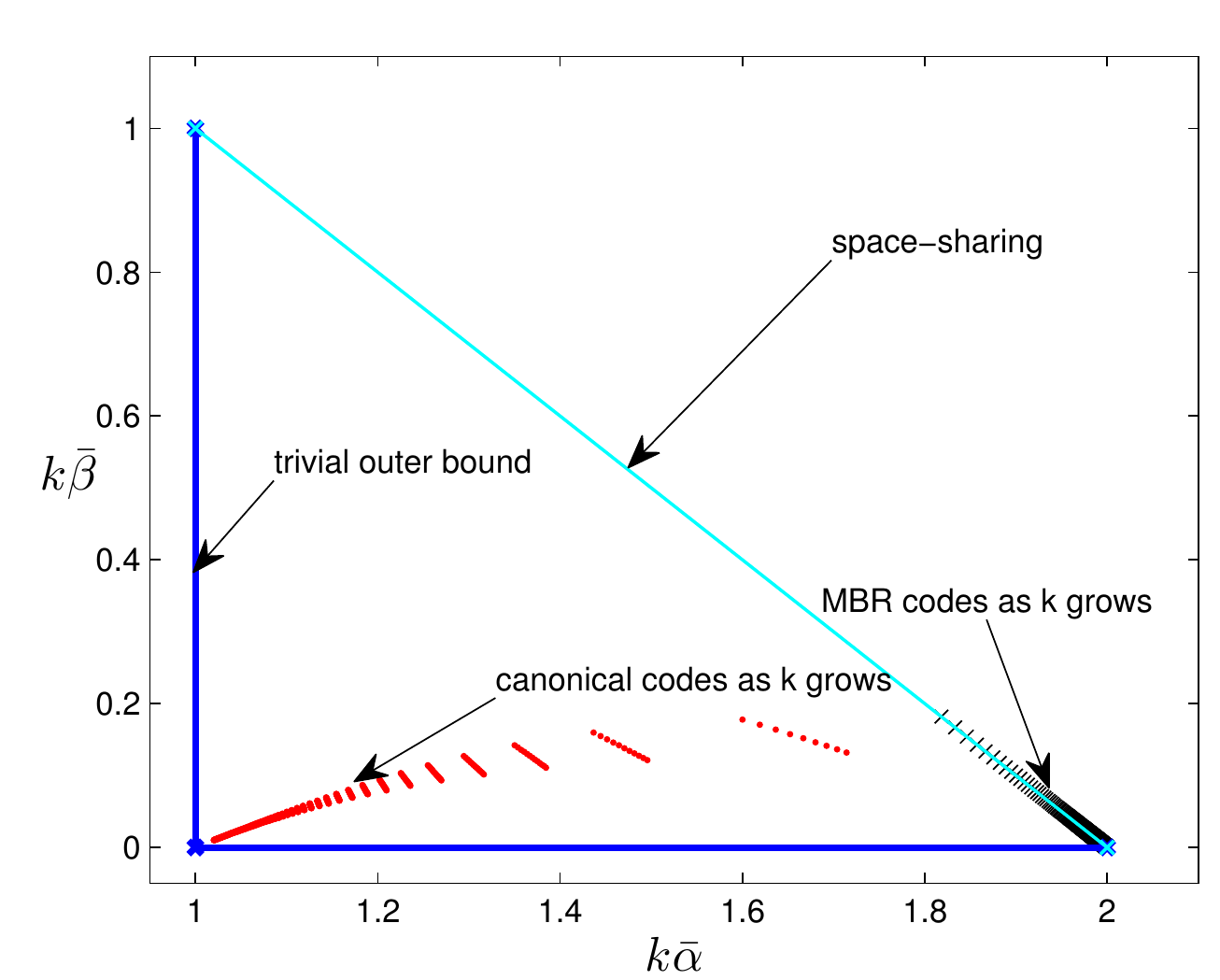}
  \caption{The asymptotic tradeoff $\mathcal{R}_\infty$.\label{fig:asympto}}
\end{figure}

In Fig. \ref{fig:asympto} we plot the trivial outer bound for $\mathcal{R}_\infty$, the MBR point cloud as $k\rightarrow \infty$, the space-sharing line, and the tradeoff points achieved by the canonical codes using the parameters given in the proof above as $k\rightarrow \infty$. It should be noted that taking any sequence of $r\sim O(k^{\delta})$ will result in the same asymptote given above, as long as $\delta\in(0,1)$. This asymptote only captures the first order behavior, and the result implies that for this case, there is in fact no asymptotic difference between functional-repair and exact-repair.

\subsection{Property of Uniform Rank Accumulation} \label{sec:ura}

Thus far, we have described the canonical code in terms of the structure of the codeword.  We now turn to a generator matrix viewpoint of the code, as the code is linear.  To obtain a generator matrix, one needs to vectorize the code array, thus replace it with a vector of size $n \alpha=rN$.  The generator matrix then describes the linear relation between the $M_c = (r-m)N$ input symbols of the canonical code ${\cal C}_{\text{can}}$ and the $n\alpha$ output symbols.  Thus the generator matrix is of size $(M_c \times n\alpha)$.  

The ordering of columns within the generator matrix is clearly dependent upon the manner in which vectorization of the code matrix takes place.   We will present two vectorizations and hence, two generator matrices:
\ben
\item From the distributed storage network point of view, each row of the $(n \times \alpha)$ code matrix corresponds to a node in distributed storage network.   Thus a natural vectorization is one in which the $n\alpha$ code symbols are ordered such that the first $\alpha$ symbols correspond to the elements of the first row vector (in left-to-right order), of the code matrix, the second $\alpha$ symbols correspond in order, to the elements of the second row vector, etc..  Thus, under this vectorization, the first $\alpha$ columns of the generator matrix correspond to the first row vector of the code array and so on.  We will refer to this as the {\em node-wise vectorization} of the code.  We will use $\vec{G}$ to denote the generator matrix of the canonical code ${\cal C}_{\text{can}}$ under this vectorization.  Each set of columns of $\vec{G}$ corresponding to a node of the codeword array, will be referred to as a {\em thick column}. In other words, the code symbols associated to the $i$-th thick column of the generator matrix are the code symbols stored in the $i$-th storage node. In this context, we will refer to a single column of $\vec{G}$ as a {\em thin column}. 
\item The code symbols in the code array of the canonical code ${\cal C}_{\text{can}}$ can be vectorized in a second manner such that the resultant code vector is the serial concatenation of the $N$ MDS codewords $\{\vec{c}_{i}\}$, each associated with a distinct message vector $\vec{u}_{i}$. We will refer to this as the {\em parity-group-wise vectorization} of the code. Let $G_{\text{b-d}}$ denote the associated generator matrix of ${\cal C}_{\text{can}}$.  Clearly, $G_{\text{b-d}}$ has a block-diagonal structure:
\bea \label{eq:Gbd}
\vec{G_{\text{b-d}}} & = & \left[
\begin{array}{cccc}
 \vec{\gmds} &   &   & \\
  & \vec{\gmds}  &   & \\
  &   & \ddots  & \\
  & & & \vec{\gmds}
\end{array}
\right].
\eea
Here \vec{\gmds} denotes the generator matrix of the $[r, r-m]$-MDS code denoted by \cmds. It follows that the columns of $\vec{G_{\text{b-d}}}$ associated with code symbols belonging to distinct parity groups span subspaces that are linearly independent.  Also, any collection of $(r-m)$ columns of $\vec{G_{\text{b-d}}}$ associated with the same parity group are linearly independent.  
\een

We will now establish that the matrix $\vec{G}$ has the following {\em $t$-uniform rank-accumulation property ($t$-URA)}: if one selects a set $T$ of $t$ thick columns drawn from amongst the $n$ thick columns of $\vec{G}$, then the rank of the submatrix $\vec{G}|_T$ of $\vec{G}$ is independent of the choice of $T$; we call a code satisfying this property a $t$-URA code. Hence the rank of $\vec{G}|_T$ may be denoted as $\rho_t$, indicating that it does not depend on the specific choice of $T$ of cardinality $t$. If a code is $t$-URA for all $t=1,2,\ldots,n$, then we say the code satisfies the universal-URA property, or that it is a universal-URA code. 
 
The value of $\rho_t$ can be determined from how the collection of thin columns in $T$ intersect with the blocks of $\vec{G}_{\text{b-d}}$. More specifically, due to the linear independence structure of columns of $\vec{G_{\text{b-d}}}$, we only need to count the total number of linear independent columns in $\vec{G_{\text{b-d}}}$ that correspond to the thin columns of $\vec{G}|_T$. The values of $\rho_t$ for DCBDs and BIBDs can be derived as follows:
\begin{itemize}
\item For the codes based on DCBDs, within each parity group, the number of columns chosen can range from $0$ to $r$. If the intersection is of size $p$, the rank accumulated is $\min\{p,r-m\}$, and thus it follows that 
\bea 
\label{eq:rho}
\rho_t \ = \nu \sum_{\substack{p=\max\{1, \\ r-(n-t) \}}}^{\min\{t,r\}} {t \choose p} {n-t \choose r-p} \min\{p,r-m\} . 
\eea  
These codes satisfy the universal-URA property, and it can be verified that $\rho_t = N(r-m) = M_c$ for $t \geq k$. 
\item For the canonical codes based on restricted Steiner systems and BIBDs, the $t$-URA property holds when $t=n$, $n-1$ or $n-2$, but in general not for other values. It is straightforward to verify that
\begin{align}
\rho_{n}=\rho_{n-1}=N(r-1)=\frac{\lambda n(n-1)}{r},
\end{align} 
and  
\begin{align}
\rho_{n-2}=N(m-1)-\lambda=\frac{\lambda n(n-1)}{r}-\lambda,
\end{align} 
because the pair of indices of the lost nodes appears in exactly $\lambda$ blocks in $S_{\lambda}(r,n)$, and for each of the involved parity group, we only collect $(r-2)$ columns in  $\vec{G}_{\text{b-d}}$, instead of $(r-1)$.
\end{itemize}

\section{Code Constructions for $d>k$}
\label{sec:general}

In this section, we first describe an explicit code construction for $[n,n-2,n-1]$ code based on restricted Steiner systems $S(r,n)$. This construction however only applies to the case when a restricted Steiner system exists for such $n$, and as aforementioned, Steiner systems may not exist for all $(r,n)$ pairs. Then construction using DCBDs for general values $[n,k,d]$ are presented based on linearized polynomials. The alphabet size of the second class of codes can be quite large, and we show that it can be reduced significantly. The performance of the code is then discussed. 

\subsection{Constructions Based on Restricted Steiner Systems and BIBDs for $[n,k=n-2,d=n-1]$}
\label{sec:SteinerConstruction}

Given a restricted Steiner system $S(r,n)$, a canonical code can be constructed with  $[n,k,d=k=n-1]$ as shown in the previous section. Next we construct a code with $[n,k=n-2,d=n-1]$ by using an additional encoding step. The alphabet can be chosen to be $\mathbb{F}_q$, where $q\geq r$, and the number of data symbols is $M=N(r-1)-1$. Let the data symbols be written in an $(r-1)\times N$ matrix except the bottom-right entry $u_{N,r-1}$, which is parity symbol given the following value
\begin{align}
\label{eqn:longMDS}
u_{N,r-1}=\sum_{j=1}^{r-2}\phi_j\sum_{i=1}^{N} u_{i,j}+\phi_{r-1}\sum_{i=1}^{N-1}u_{i,r-1},
\end{align}
where $\phi_j$\rq{}s are distinct non-zero values in $\mathbb{F}_q$, and additionally  $\phi_{j}+1\neq 0$ for $1\leq j\leq r-2$.
With this new $\vec{U}$ data matrix, we then apply the canonical code encoding procedure to produce the $n\times \gamma$ code array. The repair procedure with 
$d=n-1$ helper nodes is precisely the same as in the previous section, and thus  for this code
\begin{align}
\alpha=\frac{n-1}{r-1},\quad\beta=1,\quad M=(r-1)N-1=\frac{n(n-1)}{r}-1.
\end{align}
Note that these parameters are all integers for a valid Steiner system.  Next we show that this code indeed can recover all the data symbols using any $k=n-2$ nodes. Recall that for a restricted Steiner system, any pair of nodes appears only once in the block design, and thus only a single parity group loses two symbols when any two nodes have failed. For parity groups losing only one symbol or less, all the symbols within them can be recovered, and thus only the parity group that loses exactly two symbols need to be considered. Taking this fact into consideration, the following cases need to be considered:
\begin{enumerate} 
\item The $i$-th parity group $\vec{c_i}$, $i<N$, loses two symbols, one is a data symbol $u_{i,j}$ where $j<r$, and the other is the parity symbol $c_{i,r}$. The only missing data symbol $u_{i,j}$ can be obtained by eliminating in (\ref{eqn:longMDS}) all the other data symbols. 

\item The $i$-th parity  $\vec{c_i}$, $i<N$, loses two symbols, which are both data symbols $u_{i,j_1}$ and $u_{i,j_2}$, where $j_1<j_2<r$. Since $u_{N,r-1}$ is available, by eliminating all other other data symbols, we obtain the value of $\phi_{j_1}u_{i,j_1}+\phi_{j_2}u_{i,j_2}$. Next by eliminating all other data symbols in $c_{i,r}=\sum_{j=1}^{r-1}u_{i,j}$, we obtain the value of $u_{i,j_1}+u_{i,j_2}$. Since $\phi_{j_1}\neq \phi_{j_2}$ and they are both non-zero, $u_{i,j_1}$ and $u_{i,j_2}$ can be solved using these two equations.

\item Parity group $\vec{c_N}$ loses two symbols, which are $u_{N,r-1}$ and $c_{N,r}$. This case is trivial since all data symbols have been directly recovered. 

\item Parity group $\vec{c_N}$ loses two symbols, which are the parity symbols $c_{N,r}$ and a data symbol $u_{N,j}$, $1\leq j\leq r-2$. By eliminating the other data symbols in $u_{N,r-1}$ using (\ref{eqn:longMDS}), we obtain $u_{N,j}$. 

\item Parity group $\vec{c_N}$ loses two symbols, which are $u_{N,r-1}$ and data symbol $u_{N,j}$,  $1\leq j\leq r-2$. Note that 
\begin{align}
c_{N,r}=\sum_{j=1}^{r-1}u_{N,j}=\sum_{j=1}^{r-2}\phi_j\sum_{i=1}^{N} u_{i,j}+\phi_{r-1}\sum_{i=1}^{N-1}u_{i,r-1}+\sum_{j=1}^{r-2}u_{N,j}.
\end{align}
By eliminating all the other data symbols from $c_{N,r}$, we obtain the value of $(\phi_{j}+1)u_{N.j}$. Since $\phi_{j}+1\neq 0$ for $1\leq j\leq r-2$, the only missing data symbol $u_{N,i}$ can be obtained. 
\end{enumerate}

\begin{figure}[tcb]
  \centering
  \includegraphics[width=10cm]{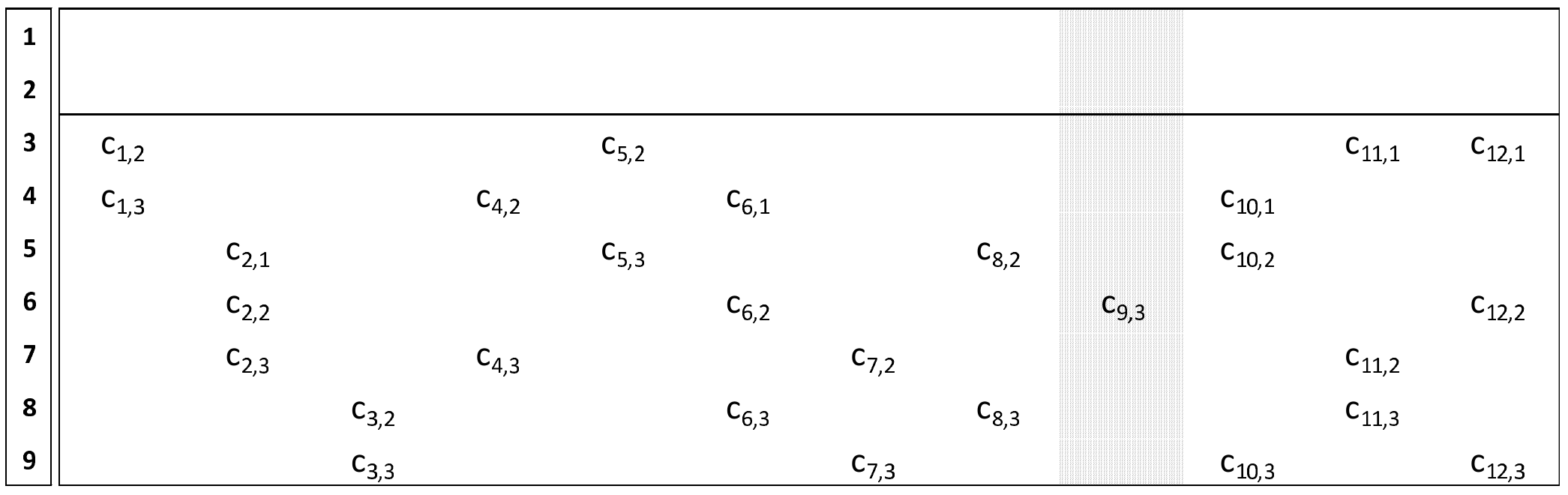}
  \caption{\label{fig:988exampleE}The $(9,7,8)$ code based using the canonical code in Fig. \ref{fig:cutset}. When node-$1$ and node-$2$ have failed, except parity group $9$, all other parity groups at least have two symbols remaining and thus can be completely recovered. To recover the symbols $c_{9,1}$ and $c_{9,2}$, the parity symbols $c_{9,3}$ and $c_{12,2}$ provide sufficient information.}
\end{figure}

There are essentially two MDS codes in this construction: the first code (referred to as the long MDS code) in the construction is an $[M+1,M]$ systematic MDS code whose parity symbol is specified by (\ref{eqn:longMDS}), and the component code (referred to as the short code) is an $[r,r-1]$ systematic MDS code. The key is to jointly design the two codes, and thus they are useful together. In the above construction, this is accomplished through the coefficients of the parity symbols. It should be noted that the coefficients in forming the parity symbols are not unique, and we have only given a convenient choice here.  

There is an inherent connection between the construction given above and the URA property of the canonical codes. Let us denote the generator matrix of the long MDS code as $\vec{G_L}$, which is  of size $M\times (M+1)$, and the generator matrix $\vec{G}$ of the canonical code in its node-wise vectorization form is of size $(M+1)\times  n\alpha$. Because of the encoding procedure, the code we eventually obtain has generator matrix $\vec{G_L}\cdot \vec{G}$ which is of size $M\times n\alpha$. To guarantee all data symbols recoverable from any $n-2$ nodes, we need the submatrix of $\vec{G_L}\cdot \vec{G}$ formed by collecting any $(n-2)$ thick columns to have rank at least $M$, which is equivalent to having $\vec{G_L}\cdot \vec{G}|_T$ to have rank at least $M$, for any $T\subset I_n$ and $|T|=n-2$. The $(n-2)$-URA property of the canonical codes implies that that $\vec{G}|_T$ has rank $\rho_{n-2}$, and thus $\rho_{n-2}$ is an upper bound on $M$; our code construction above is indeed able to achieve $M=\rho_{n-2}$.

To generalize the above construction and allow canonical codes based on BIBDs, we need to carefully choose the coding coefficients in the long MDS code such that the upper bound $M\leq \rho_{n-2}$ can be achieved with equality. In the following, an explicit construction based on rank-metric code is provided in the context of canonical codes using DCBDs, which can also be used with canonical codes based on BIBDs, and it leads to
\begin{align}
\alpha=\frac{\lambda(n-1)}{r-1},\quad\beta=\lambda,\quad M=(r-1)N-\lambda=\frac{\lambda n(n-1)}{r}-\lambda.
\end{align}

\subsection{A Construction Based on DCBDs for General $[n,k,d]$}

For the more general settings of $[n,k,d]$ that are not limited to $[n,k=n-2,d=n-1]$ (or when the corresponding restricted Steiner system does not exist), the coding coefficients in the long MDS code need to chosen carefully such that the upper bound $M\leq\rho_k$ of the canonical codes can be achieved with equality. The construction presented next utilizes  Gabidulin codes to achieve this goal.    

Let $r-m\leq k$, and choose $m=n-d$. Fix a $C_\nu(r,n)$ and the corresponding canonical code in $\mathbb{F}_q$, the number of data symbols $M$ in this new code is chosen to be equal to the upper bound $\rho_k$ in the canonical code. The $M$ message symbols $\{v_i\}_{i=1}^M$, $v_i\in \mathbb{F}_{q^\kappa}$ are first used to construct a linearized polynomial 
\bean
f(x) & = & \sum_{i=1}^{M} v_i x^{q^{i-1}},
\eean 
where $\kappa$ is any sufficiently large positive integer, and we shall provide a lower bound for its value in the sequel. 
The linearized polynomial is then evaluated at $M_c=(r-m)N$ elements $\{\theta_{i,j}\}$ of $\mathbb{F}_{q^\kappa}$, $i=1,2,\ldots,N$, $j=1,2,\ldots,r$,  which when viewed as vectors over $\mathbb{F}_q$, are linearly independent. This coding step is not systematic, however a systematic version of the code can be obtained straightforwardly by equating the data symbols as the first $\rho_k$ outputs $\left(f(\theta_1),f(\theta_2),\ldots,f(\theta_{\rho_k})\right)$, and then identifying the proper coefficients $v_i$\rq{}s; from here on we do not distinguish these two cases.

We wish to feed these $M_c$ evaluations $\{f(\theta_{i,j})\}$ into an encoder for the afore-chosen canonical code by setting the elements of the input data matrix $\vec{U}$,
\bean
u_{i,j} & = & f(\theta_{i,j}), \qquad  1 \leq i \leq N, \quad 1\leq j\leq r.
\eean
However, notice that in the original canonical code, the elements  in the data matrix input $u_{i,j}\in\mathbb{F}_q$, and the evaluations of the linearized polynomial $f(\theta_{i,j})\in\mathbb{F}_{q^\kappa}$. This discrepancy can be resolved by taking the standard convention of viewing  $\{f(\theta_{i,j})\}$  as vectors over $\mathbb{F}_q$, and apply the canonical code encoder over each of their components\footnote{Equivalently, this is the field operation in $\mathbb{F}_{q^\kappa}$ when the canonical code coefficients are viewed as in the corresponding base field $\mathbb{F}_{q}$ elements correctly embedded in the extended field.}; we use the same convention on the outputs, and thus obtain a code array of $n\times \alpha$ over $\mathbb{F}_{q^\kappa}$ through the canonical code encoding process. 
 
It is clear that the repair procedure is precisely the same as the underlying canonical code, and thus we only need to show that it is possible to recover the message symbols $\{v_i\}_{i=1}^M$ by connecting to an arbitrary set of $k$ nodes. 

\begin{prop} \label{thm:data_collection_linearized} 
By connecting to an arbitrary set of $k$ nodes, a data collector will be able to recover the message symbols $\{v_i\}_{i=1}^{M}$ in the above code.
\end{prop}

\bpf Let $\vec{G}$ denote the generator matrix of the canonical code when node-wise vectorization is employed. Observe that the entries in $\vec{G}$ belong to $\mathbb{F}_q$. 

Let $(c_1,c_2,\cdots, c_{n\alpha})$ denote the node-wise vectorized codeword of ${\cal C}$. Then we have
\bean
(c_1,c_2,\cdots,c_{n\alpha}) & = & [f(\theta_1) \ f(\theta_2) \ \cdots f(\theta_{M_c})] \cdot \vec{G} .
\eean
Using linearity of $f(\cdot)$, we can write this as
\bean
(c_1,c_2,\cdots,c_{n\alpha}) & = & f( [\theta_1 \ \theta_2 \ \cdots \theta_{M_c}] \cdot \vec{G}]) \\
& = & f(\underbrace{[\vec{x_1} \ \vec{x_2} \cdots \vec{x_{M_c}}]}_{(N \times M_c)}\cdot \vec{G}]) ,
\eean
in which $\vec{x_i} \in \mathbb{F}_q^\kappa$ is the vector representation of the element $\theta_i \in \mathbb{F}_{q^M}$, with respect to some basis 
of $\mathbb{F}_{q^\kappa}$ over $\mathbb{F}_q$. Set
\bean
\vec{X} & = & [\vec{x_1} \ \vec{x_2} \cdots \vec{x_{M_c}}].
\eean
Now let $A$ be the set of $k$ thick columns of $\vec{G}$, corresponding to the set of nodes to which the data collector is connecting to. Since $\{ \vec{x_i} \}_{i=1}^{M_c}$ are linearly independent over $\mathbb{F}_q$, it follows that 
\bea \label{eq:ura_carries_over}
\text{Rank}\left( \vec{X} \cdot \vec{G}|_A \right) & = & \text{Rank}\left( \vec{G}|_A \right) \\
& = & \rho_k \ = \ M \nonumber 
\eea
Hence there are at least $M$ linearly independent columns in the matrix product $\vec{X} \cdot \vec{G}|_A$. These columns correspond to linearly independent points of $\mathbb{F}_{q^\kappa}$ over $\mathbb{F}_q$. Thus $f\left( \vec{X} \cdot \vec{G}|_A \right)$ yields the evaluations of $f(\cdot)$ at at least $M$ linearly independent points of $\mathbb{F}_{q^\kappa}$. By Lemma ~\ref{lem:useful}, $f$ and thereby its coefficients can be uniquely identified from these $M$ evaluations. 
\epf

It is clear that the performance of the code is given by
\begin{align}
&\quad M=\rho_k \ = \nu \sum_{\substack{p=\max\{1, \\ r-(n-k) \}}}^{\min\{k,r\}} {k \choose p} {n-k \choose r-p} \min\{p,r-m\},\nonumber\\
&\alpha=\gamma=\nu{n-1 \choose r-1},\quad \beta=\frac{(r-m)\alpha}{n-m}=\frac{(r-m)\nu}{n-m}{n-1 \choose r-1}.
\label{eqn:noncanonical}
\end{align}

It should be noted that if we choose $r=2$, $m=1$, the construction reduces to the repair-by-transfer MBR code given in \cite{RashmiShah:12:1}. It is thus not surprising that the construction given here has the help-by-transfer property, since it includes the repair-by-transfer code as a special case. 

Since the canonical code exists when $q\geq r$, and $\mathbb{F}_{q^\kappa}$ must have at least $ M_c= \nu \cdot {n \choose r} (r-m)$ linearly independent elements over $\mathbb{F}_q$, we require
$\kappa  \geq  M_c$. Hence a finite field of size $r^{M_c}$ is sufficient in the above construction (exponential in $r$). We show in the next subsection that there exist constructions of significantly lower field size (linear in $r$).

For the case $[n,k=n-2,d=n-1]$, DCBD-based canonical codes can also be used even when the corresponding restricted Steiner systems exist. The advantages of the construction given in the previous subsection are that: firstly it induces smaller $\alpha$ and $\beta$ values, secondly, the required alphabet size is smaller than the one specified above (and the one shown to exist in the sequel), and lastly the coding coefficients are more explicitly specified.  

\subsection{Existence of Codes with Lower Field Size}
As aforementioned in Section \ref{sec:SteinerConstruction}, the code for the general parameters has a generator matrix in the form $\vec{G_L}\cdot\vec{G}$, where $\vec{G_L}$ is from the long MDS code, and $\vec{G}$ is from the canonical code (short MDS code), which is the node-wise vectorization version. We can alternatively consider the parity-group-wise vectorization version, which is $\vec{G_L}\cdot\vec{G}_{\text{b-d}}$. Clearly the code corresponding to $\vec{G_L}\cdot\vec{G}_{\text{b-d}}$ is a subspace of the rowspace $\vec{C}$ of $\vec{G}_{\text{b-d}}$. In other words, the dual code of $\vec{G_L}\cdot\vec{G}_{\text{b-d}}$ is a superspace of the dual $\vec{C}^\perp$ of $\vec{C}$. Suppose 
\bea \label{eq:Hbd}
\vec{H_{\text{b-d}}} & = & \left[
\begin{array}{cccc}
 \vec{\hmds} &   &   & \\
  & \vec{\hmds}  &   & \\
  &   & \ddots  & \\
  & & & \vec{\hmds}
\end{array}
\right].
\eea
is a parity-check matrix of $\vec{C}$. Here \vec{\hmds} denotes the parity-check matrix of the $[r, r-m]$-MDS code \cmds. We need to enlarge the rowspace of $\vec{H_{\text{b-d}}}$ by adding more rows to it in order to make it a parity-check matrix of the code with generator matrix $\vec{G_L}\cdot\vec{G}_{\text{b-d}}$. Let 
\bean
\vec{H} & = & \left[ \begin{array}{c} \vec{H_{\text{b-d}}} \\
					\vec{H}_1 \end{array} \right] 
\eean
be the resultant parity-check matrix. Conversely, any matrix $\vec{H}_1$ essentially specifies a subspace of the canonical code that is the rowspace of $\vec{G}_{\text{b-d}}$. For any such subspace, there always exists a matrix $\vec{G_L}$ such that the rows of $\vec{G_L}\cdot\vec{G}_{\text{b-d}}$ span the chosen subspace. Hence specifying  $\vec{G_L}$ is equivalent to specifying $\vec{H}_1$. We denote the elements of $\vec{H}_1$ as $h_{i,j}$, which are to be determined; fix an  $[r,r-m]$ MDS code in the canonical code construction, which thus implies that the matrix $\vec{H_{\text{b-d}}}$ is fixed.

For any set $T\subset I_n$ of nodes, where $|T|=k$, there are $k$ thick columns in $\vec{G_L}\cdot\vec{G}_{\text{b-d}}$ corresponding to these nodes. If and only if the submatrix formed by collecting these $k$ thick columns in $\vec{G_L}\cdot\vec{G}$ has rank $M=\rho_k$, can we recover all the $M$ data symbols from these $k$ nodes. Let us consider a submatrix $\vec{G_L}\cdot\vec{G}\rq{}|_T$, where $\vec{G\rq{}}|_T$ is formed by the following procedure: for each parity group $\vec{c}_i$, $i=1,2,\ldots,N$,
\begin{itemize}
\item When there are more than $(r-m)$ thin columns corresponding to the same parity group $\vec{c}_i$ in the $k$ thick columns, then collecting any $(r-m)$ of them;
\item Otherwise, collect all the thin columns corresponding to the remaining code symbols in this parity group.
\end{itemize}
It is clear that this results in $\rho_k$ columns. Let $S_T \subset I_{n\alpha}$ denote the indices of these $\rho_k$ thin columns. If this $\rho_k\times \rho_k$ matrix $\vec{G_L}\cdot\vec{G}\rq{}|_T$ has full rank, then all the $M$ data symbols can be recovered from the $k$ nodes. This is equivalent to having $(n\alpha - \rho_k) \times (n\alpha - \rho_k)$ submatrix $\vec{H}|_T$ of $\vec{H}$ restricted to those thin columns indexed by $I_{n\alpha} \setminus S_T$ to have full rank. This requires the determinant of  $\vec{H}|_T$ be not zero, and we write the determinant as a polynomial $f_{T} ( \{ h_{i,j}\mid i \in I_{n\alpha - \rho_k - Nm}, j \in I_{n\alpha}\})$.  Now, define
\bea \label{eq:det_poly}
p(\{ h_{i,j} \}) & = & \prod_{T\subset I_n:|T|=k} f_{T} ( \{ h_{i,j}\mid i \in I_{n\alpha - \rho_k - Nm}, j \in I_{n\alpha}\}). 
\eea
If there exists an assignment for $\{ h_{i,j} \}$ such that the polynomial $p(\cdot)$ evaluates to a non-zero value, then such an assignment will yield a $\vec{G_L}$ that ensures the required data-collection property. We make use of the following lemma from \cite{Alo} at this point.

\begin{lemma} \cite{Alo} \label{lem:nullstellansatz} ({\em Combinatorial Nullstellansatz}) Let $\mathbb{F}$ be a field, and let $f = f(x_1 , \cdots , x_n)$ be a polynomial in $\mathbb{F}[x_1, \cdots , x_n ]$. Suppose the degree $\text{deg}(f)$ of $f$ is expressible in the form $\sum_{i=1}^{n} t_i$, where each $t_i$ is a non-negative integer and suppose that the coefficient of the monomial term $\prod_{i=1}^{n} x_i^{t_i}$  is nonzero. Then if $S_1 ,\ldots , S_n$ are  subsets of $\mathbb{F}$ with sizes $|S_i|$ satisfying $|S_i| > t_i$, then there exist elements $s_1 \in S_1, s_2 \in S_2 \ldots , s_n \in S_n$ such that $f(s_1, s_2 , \cdots , s_n ) \neq 0$.
\end{lemma}

The condition that coefficient of the monomial term $\prod_{i=1}^{n} x_i^{t_i}$  is nonzero is equivalent to requiring $f = f(x_1 , \cdots , x_n)$ is not identically zero. We note that $f_{T} ( \{ h_{i,j}\mid i \in I_{n\alpha - \rho_k - Nm}, j \in I_{n\alpha}\})$ is indeed not identically zero, because the code construction given in the previous subsection essentially provides a non-zero assignment. 

Since the degree of any indeterminate in each of $f_{T} ( \{ h_{i,j}\mid i \in I_{n\alpha - \rho_k - Nm}, j \in I_{n\alpha}\})$ is $1$, the maximum among the degrees of a single indeterminate in $p(.)$ is upper bounded by ${n \choose k}$. Hence by Lemma~\ref{lem:nullstellansatz}, it is possible to find a suitable assignment for $\{ h_{ij} \}$, if the entries are picked from a finite field of size $ \geq {n \choose k}$. Thus we have proved the following proposition.

\begin{prop} \label{prop:low_fieldsize_existence} An $[n,k,d>k]$ non-canonical regenerating code exists over $\mathbb{F}_{q}$ with $q \geq {n \choose k}$.
\end{prop}

It should be noted that to find such a code in the given alphabet is not trivial, and a possible approach is to randomly assign the coefficients and then check whether all the full rank conditions are satisfied. 
\subsection{Performance Assessment of the General Codes}

There does not seem to be any simplification of (\ref{eqn:noncanonical})  for specific $[n,k,d]$ parameters. We provide a few examples to illustrate the performance of the codes. In Fig. \ref{fig:cutset}, we have plotted the performance of the proposed codes for the case of $[n,k,d]=[9,7,8]$, together with the cut-set bound and space-sharing line. There are two values for parameter $r=3$ or $r=4$ that yield tradeoffs below the space-sharing line; the proposed code also achieves the MBR point. Here the code for $r=3$ is based on Steiner systems, while for $r\geq 4$, the DCBD based design is used. The operating point $(\bar{\alpha},\bar{\beta})\approx(0.15,0.075)$ is also worth noting, because although it is not as good as the MSR point, and in fact it is worse than the space-sharing line, the penalty is surprisingly small. This suggests that the proposed codes may even be a good albeit not optimal choice to replace an MSR code. 

\begin{figure}[tcb]
  \centering
  \includegraphics[width=14cm]{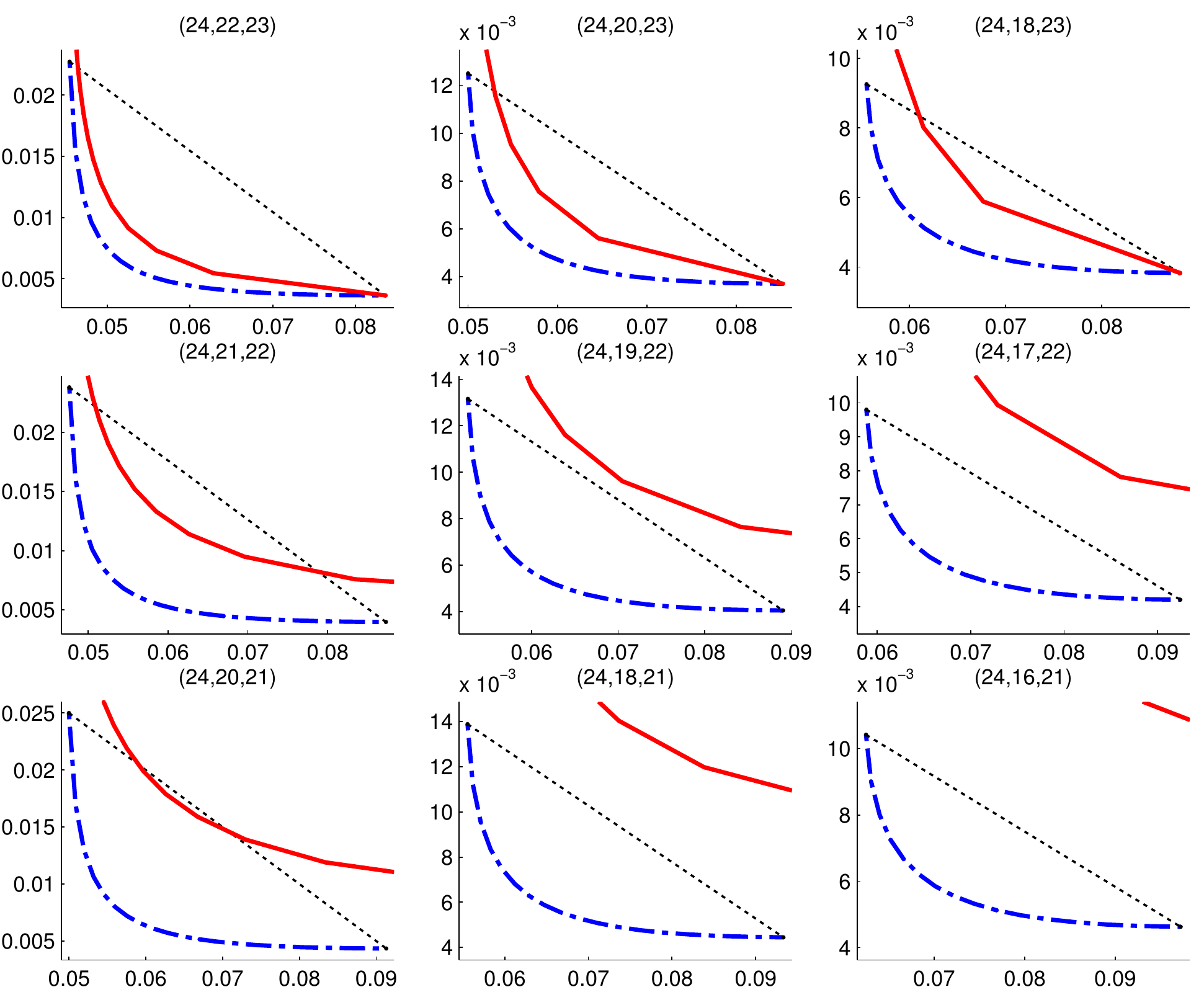}
  \caption{$\bar{\beta}$ vs $\bar{\alpha}$ for different $(k,d)$ parameters when $n=24$.\label{fig:example2} The dashed blue lines are the cut-set bounds, the dotted black lines are the space-sharing lines, and the red solid lines are the tradeoff achieved by the proposed codes. }
\end{figure}

In Fig. \ref{fig:example2} we plot the performance of codes for different parameters $(k,d)$ when $n=24$. It can be seen that when $d=n-1=23$, the performance is the most competitive, and often superior to the space-sharing line. As $d$ value decreases, the method become less effective in terms of its $(\bar{\alpha},\bar{\beta})$, and becomes worse than the space-sharing line. For the same $d$ value, the code is most effective when $k$ is large, and becomes less so as $k$ value decreases.

We can also consider the asymptotic performance of the code, however the derivation and result are almost identical to the canonical codes in the asymptotic regime we are considering ({\em i.e.,} asymptotically optimal in the sense that it achieves the complete $\mathcal{R}_\infty$), and thus we leave this simple exercise to interested readers. Another important asymptote is to keep the ratio of $k$ and $n$ constant, and letting $n\rightarrow \infty$. However, in this case, the proposed codes are not optimal asymptotically, and such an analysis does not yield further useful insight beyond the example cases shown above.

\section{Conclusion}
\label{sec:conclusion}

A new construction for $[n,k,d]$ exact-repair regenerating codes is proposed by combining embedded error correction and block designs. The resultant codes have the desirable \lq\lq{}help-by-transfer\rq\rq{} property where the nodes participating in the repair simply send certain stored data without performing any computation. We show that the proposed code is able to achieve performance better than the space-sharing between an MSR code and an MBR code for some parameters, and furthermore, the proposed construction can achieve a non-trivial tradeoff point on the functional repair tradeoff, and is in fact asymptotically optimal while the space-sharing scheme is suboptimal. For the case of $d=n-1$ and $k=n-2$, an explicit construction is given in a finite field $\mathbb{F}_q$ where $q$ is greater or equal to the block size in the combinatorial block designs. For more general  $(d,k)$ parameters, a construction based on linearized polynomial is given, and it is further shown that there exist codes with significantly smaller alphabet sizes. 

\appendix[Proof of Proposition \ref{prop:repetition}]



\begin{figure}[ht]
\centering
 \begin{center}
\includegraphics[width=10cm]{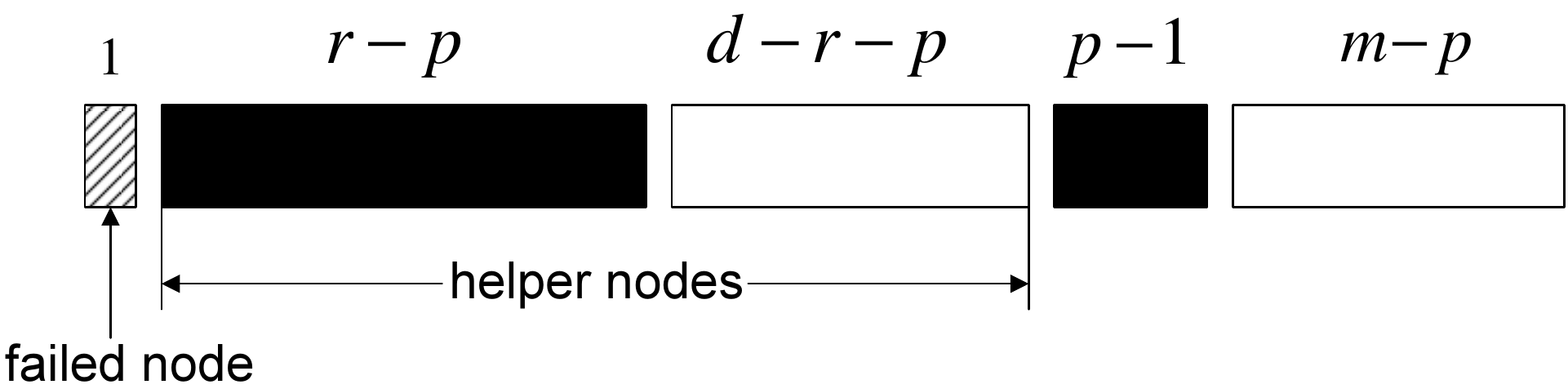} 
\end{center}
\caption{A repair situation associated to a given parameter $p$.} \label{fig:repetition_factor} 
\end{figure}

Without loss of generality, we assume that the first node has failed (see Fig.~\ref{fig:repetition_factor}) and that nodes $2$ through $(d+1)$ are the helper nodes.  
Let us focus on those blocks that contain the integer $1$ as an element.   The number of elements within such a block, that are contained amongst the helper nodes can range from $(r-m)$ to $(r-1)$.  We further focus on the blocks for which the number of elements contained amongst the helper nodes equals $(r-p)$, for a fixed value of $p$, where $1\leq p\leq m$. Denote the collection of such blocks as $L_p, 1 \leq p \leq m$. The size of $L_p$ is given by 
\bean
|L_p| & = & \nu{d \choose r-p} {m-1 \choose p-1} .
\eean
For each block in $L_p$, consider its intersection with the helper node set $I_{d+1}\setminus \{1\}$, and denote the collection of all distinct such sub-blocks as $J_p$, $1 \leq p \leq m$. The cardinality of $J_{p}$ is given by,
\bean
|J_{p}| & = & {d \choose r-p}.
\eean

A block in $J_{p}$ can equivalently be viewed as a binary vector of length $d$ and Hamming weight $(r-p)$ where the $(r-p)$ locations of the $1$s correspond to these elements in the block. Thus the set $J_{p}$ can equivalently be mapped into a $({d \choose r-p}\times d)$-binary array $P$, with each of its row vector mapping to an element in $J_p$. Let $M_i, 1 \leq i \leq {d \choose r-p}$ be the support of the $i$-th row of $P$. In any given repair strategy, each block will require to communicate $(r-m)$ symbols to the failed node, to enable repair of the failed node. Thus a repair strategy within $J_{p}$ can be described by allocating 
\bean
R_i \subseteq M_i, \ |R_i| = (r-m) 
\eean
for every $1 \leq i \leq {d \choose r-p}$. If the number of elements in a column of $P$, that belong to $R_i$ for some $i$ is equal to the same value irrespective of the choice of the column, then we refer to such a pattern of allocation for $P$ a {\em uniform allocation pattern}. Clearly a uniform allocation pattern ensures uniform download from every helper node while repairing the failed node. Let $Q$ be a binary matrix formed by stacking $P$ vertically $\mu$ times. Here $\mu$ is referred to as the repetition number. Let $M'_i, 1 \leq i \leq \mu{d \choose r-p}$ be the support of the $i$-th row of $Q$. Suppose we can identify 
\bean
R'_i \subseteq M'_i, \ |R'_i| = (r-m) 
\eean
such that the number of elements in a column of $Q$, that belong to $R'_i$ for some $i$ is equal to the same value irrespective of the choice of the column. Then we say that the repetition number $\mu$ allows a uniform allocation pattern for $P$.

In what follows, we will identify a repetition number $\mu_p$ for $P$ that allows uniform allocation. We will verify that $\mu_p \mid \nu {m-1 \choose p-1}$. Then it is clear that a repair strategy permitting uniform download from every helper node exists within the blocks of $L_p$. Since this holds true for an arbitrary value of $p$, it follows that there exists a repair strategy ensuring uniform download from each of the helper nodes.

We consider allocation for $P$ in two cases.

{\em Case 1:} $\theta_p = 1$

For any row vector ${\bf v}$ of $P$, let us call the set of all vectors that can be obtained through cyclic shifts of ${\bf v}$, the orbit of ${\bf v}$. The set $J_p$ can be partitioned into such orbits. When $\theta_p = 1$, it can be shown that all orbits are of size $d$. Consider one such orbit, and let the $(d \times d)$ submatrix $P_1$ of $P$ be the matrix formed of the vectors in the orbit arranged in such a way that the $i$-th row of $P_1$, $ 0 \leq i \leq (d-1)$ is the $i$-th periodic shift of the first row. For each $i, 0 \leq i \leq (d-1)$, we proceed to identify a subset $R_{1i}$ of the support of the $i$-th row of $P_1$. Let $M_1 \subset [d]$ be the support of the first row of $P_1$, and let $R_{10} \subseteq M_1$ be such that $|R_{10}|=(r-m)$. Let us define $R_{1i}, 0 \leq i \leq (d-1)$ as the $i$-th periodic shift of $R_{10}$. It is straightforward to see that the above choice of $\{ R_{1i} \}$ results in an uniform allocation pattern for $P_1$. The same strategy can be adopted for every orbit in $J_p$. Thus in this case of $\theta_p = 1$, the repetition factor $\mu_p = 1$ is sufficient.

{\em Case 2:} $\theta_p \neq 1$

In this case also, $J_p$ can be partitioned into orbits. Let us focus our attention to a submatrix $P_1$ of $P$ formed of the vectors in a fixed orbit. Unlike the previous case, the chosen orbit need not be of size $d$. However it can be shown that it will be of size 
\bean
\left(\frac{d}{\theta_p}\right)s & =: & \omega_{ps} 
\eean
for some $s$ such that $s \mid \theta_p$. Thus $P_1$ is a $(\omega_{ps} \times d)$ binary matrix such that the $i$-th row of $P_1$, $ 0 \leq i \leq (\omega_{ps}-1)$ is the $i$-th periodic shift of the first row. Let
\bean
\lambda_{ps} & := & \frac{d}{\omega_{ps}(\frac{d}{\omega_{ps}},r-m)_{\text{gcd}}} \ = \ \frac{\left(\frac{\theta_p}{s}\right)}{\left(\frac{\theta_p}{s}, r-m\right)_{\text{gcd}}}.
\eean
The integer $\lambda_{ps}$ is chosen as the smallest number such that $\frac{d}{\lambda_{ps}\omega_{ps}} \mid (r-m)$. Next, consider the $(\omega_{ps}\lambda_{ps} \times d)$-matrix $Q$ formed by stacking $P_1$ vertically $\lambda_{ps}$ times. It shall be noted that the matrix $Q$ has the property that its $i$-th row $0 \leq i \leq \omega_{ps}\lambda_{ps} - 1$ is the $i$-th periodic shift of its first row. The matrix $Q$ can be written as
\bean
Q & = & [Q_1 \mid Q_2 \mid \ldots \mid Q_{(\frac{d}{\omega_{ps}\lambda_{ps}})}],
\eean
where $Q_j, 1 \leq j \leq \frac{d}{\omega_{ps}\lambda_{ps}}$ is a square matrix of dimension  $\omega_{ps}\lambda_{ps}$. It can be seen that each of $\{Q_j\}$ satisfies the following properties:
\bit
\item The Hamming weight of every row equals $\frac{(r-p)\omega_{ps}\lambda_{ps}}{d}$;
\item The $i$-th row, $0 \leq i \leq \omega_{ps}\lambda_{ps}-1$ is the $i$-th periodic shift of the first row.
\eit
Let us now focus our attention on $Q_1$, and we will describe a uniform allocation pattern for $Q_1$. For each $i, 0 \leq i \leq (\omega_{ps}\lambda_{ps}-1)$, we proceed to identify a subset $R_{1i}$ of the support of the $i$-th row of $Q_1$. Let $M_1 \subset [\omega_{ps}\lambda_{ps}]$ be the support of the first row of $Q_1$, and let $R_{10} \subseteq M_1$ be such that $|R_{10}|=\frac{(r-m)\omega_{ps}\lambda_{ps}}{d}$.  Let us define $R_{1i}, 0 \leq i \leq (\omega_{ps}\lambda_{ps}-1)$ as the $i$-th periodic shift of $R_{10}$. It is not hard to see that the above choice of $\{ R_{1i} \}$ results in a uniform allocation pattern for $Q_1$. The same strategy can be adopted for $Q_j, \ 2 \leq j \leq \left(\frac{d}{\omega_{ps}\lambda_{ps}}\right)$, permitting a uniform allocation for $P_1$. Thus the repetition number of $\lambda_{ps}$ ensures uniform allocation for $P_1$, an orbit within $J_p$.

It still remains to determine a repetition number that will ensure uniform allocation for $P$. It can be shown that for every $s \mid \theta_p$, $J_p$ contains an orbit of size
\bean
\left(\frac{d}{\theta_p}\right)s & =: & \omega_{ps}  .
\eean
For every such orbit, we have already shown that a repetiton factor of 
\bean
\frac{\left(\frac{\theta_p}{s}\right)}{\left(\frac{\theta_p}{s}, r-m\right)_{\text{gcd}}}
\eean
will ensure uniform allocation within the orbit. Hence \bean
\mu_p & = & \zeta_p
\eean
allows a uniform allocation for the entire matrix $P$. 

Next, we observe that $\nu$ is chosen as the smallest number such that
\bean
\zeta_p \mid \nu {m-1 \choose p-1} 
\eean
for every $1 \leq p \leq m$. It follows that there exists a repair strategy ensuring uniform download from each of $d$ helper nodes. This completes the proof. 
\IEEEQED

We provide two examples to illustrate the design of matrix $P$ as specified in the proof above. 

\begin{example} Suppose $d=7$, $r-p=5$, $r-m=3$. Then the binary matrix corresponding to an orbit is shown below. The bold one ${\bf 1}$ represents the allocation of symbols to be transmitted for repair. 

\vspace*{0.2in}

\begin{center}
$P_1 \ = \ $ \ \ \begin{tabular}{||c|c|c|c|c|c|c||} \hline \hline 
{\bf 1} & {\bf 1} & {\bf 1} & 1 & 1 & 0 & 0  \\ \hline 
0 & {\bf 1} & {\bf 1} & {\bf 1} & 1 & 1 & 0  \\ \hline 
0 & 0 & {\bf 1} & {\bf 1} & {\bf 1} & 1 & 1  \\ \hline 
1 & 0 & 0 & {\bf 1} & {\bf 1} & {\bf 1} & 1  \\ \hline 
1 & 1 & 0 & 0 & {\bf 1} & {\bf 1} & {\bf 1}  \\ \hline 
{\bf 1} & 1 & 1 & 0 & 0 & {\bf 1} & {\bf 1}  \\ \hline 
{\bf 1} & {\bf 1} & 1 & 1 & 0 & 0 & {\bf 1}  \\ \hline 
  \end{tabular} .
 \end{center} 
 \vspace*{0.2in}

\end{example}

\begin{example} Suppose $d=6$, $r-p=4$, $r-m=2$. Then the binary matrix corresponding to an orbit is shown below. The size of the orbit $\omega_{ps} = 3$. Here we obtain $\lambda_{ps}=1$. The bold one ${\bf 1}$ represents the allocation of symbols to be transmitted for repair. 

\vspace*{0.2in}

\begin{center}
$Q \ = \ $ \ \ \begin{tabular}{||c|c|c|c|c|c||} \hline \hline 
{\bf 1} & 1 & 0 & {\bf 1} & 1& 0 \\ \hline 
0 & {\bf 1} & 1 & 0 & {\bf 1}& 1 \\ \hline 
1 & 0 & {\bf 1} & 1 & 0 & {\bf 1} \\ \hline 
  \end{tabular} .
 \end{center} 
 \vspace*{0.2in}

\end{example}

\bibliographystyle{IEEEbib}

\begin{thebibliography}{}

\bibitem{hadoop}
``{Hadoop},'' http://hadoop.apache.org.

\bibitem{zaharia2010delay}
M.~Zaharia, D.~Borthakur, J.~Sen~Sarma, K.~Elmeleegy, S.~Shenker, and
  I.~Stoica, ``Delay scheduling: a simple technique for achieving locality and
  fairness in cluster scheduling,'' in \emph{ACM Eurosys}, 2010, pp. 265--278.




\bibitem{Dimakis:10}
A. G. Dimakis, P. B. Godfrey, Y. Wu, M. Wainwright and K. Ramchandran,
\newblock ``Network coding for distributed storage systems,''
\newblock {\em IEEE Trans. Information Theory}, vol. 56, no. 9, pp. 4539-4551, Sep. 2010.

\bibitem{Yeung:00}
R. Ahlswede, Ning Cai, S.-Y.R. Li, and R. W. Yeung,
\newblock ``Network information flow,''
\newblock {\em IEEE Trans. Information Theory}, vol. 46, no. 4, pp. 1204-1216, Jul. 2000.


\bibitem{Wu:10}
Y. Wu,
\newblock ``Existence and construction of capacity-achieving network codes for distributed storage,''
\newblock {\em IEEE Journal on Selected Areas in Communications}, vol. 28, no. 2, pp. 277-288, Feb. 2010.

\bibitem{Dimakis:11}
A. G. Dimakis, K. Ramchandran, Y. Wu, C. Suh,
\newblock \lq\lq{}A survey on network codes for distributed storage,\rq\rq{} 
\newblock {\em Proceedings of the IEEE}, vol. 99, no. 3, pp. 476-489, Mar. 2011.

\bibitem{RashmiShah:12:1}
N. B. Shah, K. V. Rashmi, P. V. Kumar and K. Ramchandran,
\newblock \lq\lq{}Distributed storage codes with repair-by-transfer and non-achievability of interior points on the storage-bandwidth tradeoff,\rq\rq{}
\newblock {\em IEEE Transactions on Information Theory}, vol. 58, no. 3, pp. 1837-1852, Mar. 2012. 
 
\bibitem{RashmiShah:12:2}
N. B. Shah, K. V. Rashmi, P. V. Kumar and K. Ramchandran,
\newblock \lq\lq{}Interference alignment in regenerating codes for distributed storage: necessity and code constructions,\rq\rq{}
\newblock {\em IEEE Transactions on Information Theory}, vol. 58, no. 4, pp. 2134-2158, Apr. 2012. 

\bibitem{RashmiShah:11}
K. V. Rashmi, N. B. Shah, and P. V. Kumar,
\newblock \lq\lq{}Optimal exact-regenerating codes for distributed storage at the MSR and MBR points via a product-matrix construction,\rq\rq{}
\newblock {\em IEEE Transactions on Information Theory}, vol. 57, no. 8, pp. 5227-5239, Aug. 2011. 

\bibitem{Tamo:11}
I. Tamo, Z. Wang, and J. Bruck,
\newblock  \lq\lq{}Zigzag Codes: MDS array codes with optimal rebuilding,\rq\rq{}
\newblock  {\em IEEE Transactions on Information Theory}, vol. 59, no. 3, pp. 1597-1616, Mar. 2013. 


\bibitem{Cadambe:11}
V. Cadambe, S. Jafar, H. Maleki, K. Ramchandran and C. Suh,
\newblock \lq\lq{}Asymptotic interference alignment for optimal repair of MDS codes in distributed storage,\rq\rq{}
\newblock  {\em IEEE Transactions on Information Theory}, vol. 59, no. 5, pp. 2974-2987, May 2013. 

\bibitem{PapailiopoulosDimakisCadambe:11}
D. S. Papailiopoulos, A. G. Dimakis, and V. Cadambe,
\newblock \lq\lq{}Repair optimal erasure codes through Hadamard designs,\rq\rq{}
\newblock {\em IEEE Transactions on Information Theory}, vol. 59, no. 5, pp. 3021-3037, May 2013.


\bibitem{CadambeHuang:11}
V. R. Cadambe, C. Huang, S. A. Jafar, and J. Li, 
\newblock \lq\lq{}Optimal repair of MDS codes in distributed storage via subspace interference alignment,\rq\rq{}
\newblock arXiv:1106.1250.


\bibitem{Tian:12}
C. Tian,
\newblock ``{Characterizing the rate region of the $(4,3,3)$ exact-repair regenerating codes,}" {\em IEEE Journal on Selected Areas in Communications}, vol. 32, no. 5, pp. 967-975, May 2014.


\bibitem{Papailiopoulos:12INFOCOM}
D. S. Papailiopoulos, J. Luo, A. G. Dimakis, C. Huang, and J. Li,
\newblock \lq\lq{}Simple regenerating codes: network coding for cloud storage,\rq\rq{}
\newblock in Proceedings {\em 2012 IEEE INFOCOM}, Orlando FL, Mar. 2012, pp. 2801-2805. 

\bibitem{Rouayheb:10}
S. El Rouayheb and K. Ramchandran, 
\newblock \lq\lq{}Fractional repetition codes for repair in distributed storage systems,\rq\rq{}
\newblock in Proceedings {\em 48th Annual Allerton Conference on Communication, Control and Computation}, Monticello, Sep. 2010.

\bibitem{Tian:Arxiv}
C. Tian, V. Aggarwal and V. Vaishampayan, \lq\lq{}Exact-repair regenerating codes via layered erasure correction and block designs,\rq\rq{}
\newblock in Proceedings {\em 2013 IEEE International Symposium on Information Theory}, Istanbul, Turkey, Jul. 2013, pp. 1431-1435; also Arxiv: 1302.4670. 



\bibitem{Birenjith:13}
B. Sasidharan, P. V. Kumar, \lq\lq{}High-rate regenerating codes through layering,\rq\rq{}
\newblock in Proceedings {\em 2013 IEEE International Symposium on Information Theory}, Istanbul, Turkey, Jul. 2013, pp. 1611-1615; also Arxiv: 1301.6157.

\bibitem{Wicker:book}
S. Wicker,
\newblock{\em Error control systems for digital communication and storage},
\newblock Prentice Hall, 1995.


\bibitem{LidNie}
R.~Lidl and H.~Niederreiter, \emph{Finite Fields (Encyclopedia of Mathematics
  and its Applications)}.\hskip 1em plus 0.5em minus 0.4em\relax Cambridge
  University Press, 1997.

\bibitem{Gab85}
E.~M.~Gabidulin, \lq\lq{}Theory of codes with maximum rank distance,\rq\rq{}
\newblock \emph{Probl. Peredachi Inf.}, vol. 21, no. 1, pp. 3-16, 1985.


\bibitem{Colbourn:book}
C. J. Colbourn and J. H. Dinitz,
\newblock{\em Handbook of Combinatorial Designs, Second Edition (Discrete Mathematics and Its Applications)},
\newblock Chapman and Hall/CRC, Nov. 2006.

\bibitem{Bose:39}
R. C. Bose,
\newblock \lq\lq{}On the construction of balanced incomplete block designs,\rq\rq{}
\newblock {\em Annals of Eugenics}, vol. 9, no. 4, Dec. 1939, pp. 353-399.

\bibitem{Alo}
N.~Alon, ``{Combinatorial Nullstellensatz},'' \emph{Combinatorics, Probability
  and Computing}, 1999.









\end{thebibliography}

\end{document}